\documentclass[12pt]{article}
\usepackage{epsfig}
\usepackage{color}
\usepackage{a4}
\usepackage{latexsym}
\usepackage{cite}

\usepackage{pslatex}
\usepackage[latin1]{inputenc}
\usepackage[T1]{fontenc}

\usepackage{color,colordvi}
\def\colour4colour#1{\Blue{#1}}
\newcommand{\colourcolour}[1]{{\color{blue}{#1}}}

\newcommand{\hspn}{{\hspace{-4mm}}}
\newcommand{\hspp}{{\hspace{4mm}}}
\newcommand{\beq}{\begin{equation}}
\newcommand{\eeq}{\end{equation}}
\newcommand{\bea}{\begin{eqnarray}}
\newcommand{\eea}{\end{eqnarray}}
\newcommand{\nn}{\nonumber}
\newcommand{\MSb}{$\overline{\mbox{MS}}$}
\newcommand{\ra}{\rightarrow}

\newcommand{\als}{\alpha_{\rm s}}
\newcommand{\ars}{a_{\rm s}}

\newcommand{\ep}{\varepsilon}

\newcommand{\mufs}{\mu_{\!\!\,f}^{\,2}}

\newcommand{\gqqns}[1]{\gamma_{\,\rm ns}^{\,(#1)}}
\newcommand{\gqqps}[1]{\gamma_{\,\rm ps}^{\,(#1)}}

\newcommand{\gqg}[1]{\gamma_{\,\rm qg}^{\,(#1)}}
\newcommand{\ggq}[1]{\gamma_{\,\rm gq}^{\,(#1)}}
\newcommand{\ggg}[1]{\gamma_{\,\rm gg}^{\,(#1)}}

\newcommand{\Pns}[1]{P_{\rm ns}^{\,(#1)}}
\newcommand{\Pqqps}[1]{P_{\rm ps}^{\,(#1)}}

\newcommand{\Pqg}[1]{P_{\rm qg}^{\,(#1)}}
\newcommand{\Pgq}[1]{P_{\rm gq}^{\,(#1)}}
\newcommand{\Pgg}[1]{P_{\rm gg}^{\,(#1)}}

\begin{document}
\setlength{\parskip}{0.25cm}
\setlength{\baselineskip}{0.55cm}

\def\z#1{{\zeta_{#1}}}
\def\zss{\zeta_2^{\,2}}

\def\ca{{C^{}_A}}
\def\cas{{C^{\,2}_A}}
\def\cat{{C^{\,3}_A}}
\def\caf{{C^{\,4}_A}}
\def\cf{{C^{}_F}}
\def\cfs{{C^{\, 2}_F}}
\def\cft{{C^{\, 3}_F}}
\def\cff{{C^{\, 4}_F}}
\def\nf{{n^{}_{\! f}}}
\def\nfs{{n^{\,2}_{\! f}}}
\def\nft{{n^{\,3}_{\! f}}}

\def\dabctnc{{d^{\:\!abc\!}d_{abc}/n_c}}
\def\dabctnr{{ {d^{abc}d_{abc}}\over{n_c} }}

\def\DNnO{D_0}
\def\DNmO{D_{-1}}
\def\DNpO{D_{1}}
\def\DNppO{D_{2}}
\def\DNn#1{D_0^{\:#1}}
\def\DNm#1{D_{-1}^{\:#1}}
\def\DNp#1{D_1^{\:#1}}
\def\DNpp#1{D_2^{\:#1}}

\def\etaD#1{\eta^{\,#1}}
\def\nuD#1{\nu^{\,#1}}

\def\as(#1){{\alpha_{\rm s}^{\:#1}}}
\def\ar(#1){{a_{\rm s}^{\:#1}}}

\def\x1{{(1 \! - \! x)}}
\def\LntO{\ln(1\!-\!x)}
\def\Lnt(#1){\ln^{\,#1}(1\!-\!x)}

\def\muRs{{\mu_R^{\,2}}}
\def\Qs{{Q^{\, 2}}}

\def\S(#1){{{S}_{#1}}}
\def\Ss(#1,#2){{{S}_{#1,#2}}}
\def\Sss(#1,#2,#3){{{S}_{#1,#2,#3}}}
\def\Ssss(#1,#2,#3,#4){{{S}_{#1,#2,#3,#4}}}
\def\Sssss(#1,#2,#3,#4,#5){{{S}_{#1,#2,#3,#4,#5}}}

\newcommand{\pqq}[1]{p_{\rm qq}(#1)}
\newcommand{\pgg}[1]{p_{\rm gg}(#1)}
\def\pqg(#1){p_{\rm qg}(#1)}
\def\pgq(#1){p_{\rm gq}(#1)}

\def\frct#1#2{\mbox{\large{$\frac{#1}{#2}$}}}

\def\H(#1){{\rm{H}}_{#1}}
\def\Hh(#1,#2){{\rm{H}}_{#1,#2}}
\def\Hhh(#1,#2,#3){{\rm{H}}_{#1,#2,#3}}
\def\Hhhh(#1,#2,#3,#4){{\rm{H}}_{#1,#2,#3,#4}}
\def\Hhhhh(#1,#2,#3,#4,#5){{\rm{H}}_{#1,#2,#3,#4,#5}}

\def\npqq{p_{\rm qq}}
\def\npgg{p_{\rm gg}}
\def\npqg{p_{\rm qg}}
\def\npgq{p_{\rm gq}}

\def\Pnsm(#1){P_{\rm ns}^{\,-(#1)}}

\begin{titlepage}
\noindent
LTH 1106  \hfill October 2016\\[0.5mm]
NIKHEF 2016-049 \\
\vspace{1.2cm}
\begin{center}
{\Large \bf Large-n$_{\:\!\bf f}$ Contributions to the}\\
\vspace{0.3cm}
{\Large \bf Four-Loop Splitting Functions in QCD}\\ 
\vspace{2.0cm}
{\large J. Davies$^\ast$,  A. Vogt}\\
\vspace{0.4cm}
{\it Department of Mathematical Sciences, University of Liverpool\\
\vspace{0.5mm}
Liverpool L69 3BX, United Kingdom}\\
\vspace{1.2cm}
{\large B. Ruijl$^{\:\!\ddagger}$, T. Ueda and J.A.M. Vermaseren}\\
\vspace{0.4cm}
{\it Nikhef$\,$ Theory Group \\
\vspace{0.5mm}
Science Park 105, 1098 XG Amsterdam, The Netherlands} \\
\vspace{0.3cm}
$^\ddagger$ {\it Leiden Centre of Data Science, Leiden University \\
\vspace{0.5mm}
Niels Bohrweg 1, 2333 CA Leiden, The Netherlands}\\
\vspace{2.5cm}
{\large \bf Abstract}
\vspace{-0.2cm}
\end{center}
We have computed the fourth-order $\nfs$ contributions to all three non-singlet
quark-quark splitting functions and their four $\nft$ flavour-singlet 
counterparts for the evolution of the parton distributions of hadrons in 
perturbative QCD with $\nf$ effectively massless quark flavours. The analytic 
form of these functions is presented in both Mellin $N$-space and 
momentum-fraction $x$-space; the \mbox{large-$x$} and small-$x$ limits are 
discussed.  Our results agree with all available predictions derived from 
lower-order information.  The large-$x$ limit of the quark-quark cases provides
the complete $\nfs$ part of the four-loop cusp anomalous dimension which agrees
with two recent partial computations.

\vspace{1cm}
\noindent
$^\ast$ {\it Present address: Institute for Theoretical Particle Physics, 
Karlsruhe Institute of Technology, 
\\ \hspace*{3.1cm}D-76128 Karlsruhe, Germany}
\end{titlepage}
%
%
\section{Introduction}
\label{sec:intro}
In the past years the next-to-next-to-leading order (NNLO) corrections 
in perturbative QCD have been determined for many high-energy processes, 
see Refs.~\cite{NNLO1,NNLO2,NNLO3,NNLO4,NNLO5,NNLO6,NNLO7,NNLO8}
for some recent calculations.
For processes with initial-state protons, NNLO analyses require parton 
distributions evolved with the three-loop splitting functions 
\cite{mvvPns,mvvPsg}.
In some cases also the next-to-next-to-next-to-leading order (N$^3$LO)
corrections are important, e.g., for quantities with a slow convergence of 
the perturbation series or for cases where a very high accuracy is required.
An example of the former is Higgs production at proton-proton colliders
\cite{Higgs1,Higgs2}.
An example of the latter is the determination of the strong coupling constant
$\als$ from the structure functions $F_2$ and $F_3$ in lepton-nucleon
deep-inelastic scattering (DIS), see Ref.~\cite{ABMas15},
for which the N$^3$LO coefficient functions have been obtained in Refs.\
\cite{mvvF2L,mvvF3}.
In principle N$^3$LO analyses of these processes require the four-loop 
splitting functions, although estimates of these functions via, for example, 
Pad\'{e} approximants can be sufficient in some cases such as for DIS at 
large Bjorken-$x$.

At present a direct computation of the four-loop splitting functions 
$P_{\:\!\rm ik}^{\,(3)}(x)$ appears to be too difficult.
Work on low-integer Mellin moments of these functions started ten years 
ago \cite{ChP4ns1}; until recently only the $N=2$ and $N=4$ moments had
been obtained of the quark+antiquark non-singlet splitting function 
$P_{\:\!\rm ns}^{\,(3)+}$ together with the $N=3$ result for its 
quark--antiquark counterpart $P_{\:\!\rm ns}^{\,(3)-}$ 
\cite{VelizN2,VelizN34,ChP4ns2}. 
Using {\sc Forcer} \cite{tuLL2016,FORCER}, a four-loop generalization of the 
well-known {\sc Mincer} program \cite{MINCER1,MINCER2} for the parametric 
reduction of self-energy integrals, it is now possible to derive more moments 
in the same manner as in Refs.~\cite{Mom3loop1,Mom3loop2,Mom3loop3} 
at the third order in $\als$.
So far the moments up to $N=6$ and $N=4$ have been computed, respectively, for 
the non-singlet and singlet cases \cite{avLL2016,RUVV1}, and computations up
to $N=8$ are feasible.
Further conceptual and$\:\!/$or computational developments are required, however,
in order to obtain sufficient information for the construction of approximate 
$x$-space expressions analogous to those at three loops in Ref.~\cite{NVappr}.

The situation is far more favourable for the contributions to the functions
$P_{\:\!\rm ik}^{\,(3)}(x)$ which are leading (in the singlet case) or 
leading and sub-leading (in the non-singlet case) in the number $\nf$ of 
effectively massless quark flavours. Here the harder four-loop diagram
topologies do not contribute, and {\sc Forcer} calculations above $N = 20$,
and in some cases above $N = 40$, are possible.
If suitably combined with information and expectations on the structure of
these contributions in terms of harmonic sums \cite{Hsums,BKurth}, these 
fixed-$N$ results turn out to be sufficient to find and validate the analytic 
dependence of these parts of the four-loop splitting functions on $N$,
and hence on $x$ in terms of harmonic polylogarithms \cite{Hpols}, by 
LLL-based techniques \cite{LLL,axbAlg,Calc}. This approach has been used
before, e.g.~in Refs.~\cite{VelizTrv,mvvDP2} for the three-loop transversity 
and helicity-difference splitting functions, and may be applicable to other
four-loop quantities in the future.
The present results include the $\nfs$ part of the four-loop cusp anomalous 
dimension also obtained in Refs.~\cite{HSSS16,GHKM15,GrozinLL}.

The remainder of this article is organized as follows: in Section 2 we set
up our notations and briefly discuss the diagram calculations and the LLL
analyses of the resulting integer-$N$ moments. 
The analytic results for the $\nft$ parts of $P_{\:\!\rm ik}^{\,(3)}$ and 
the $\nfs$ parts of $P_{\:\!\rm ns}^{\,(3)}$ in $N$- and $x$-space are 
presented and discussed in Sections 3 and 4.
We summarize our results in Section 5.

%
\section{Notations and calculations}
\label{sec:calc}
\setcounter{equation}{0}
 
The renormalization-group evolution equations for the dependence of the 
parton momentum distributions $f_{\rm a} = u,\, \bar{u},\, d,\, \bar{d},\, 
\ldots,\, g$ of hadrons on the mass factorization scale $\mu_{\!f}^{}$,
\beq
\label{evol}
  \frac{d}{d \ln \mufs} \: f_{\:\!\rm a\!} \left(x,\mufs \right)
  \:\:=\:\: 
  \int_x^1 \! \frac{dy}{y} \: P_{\rm ab}(y,\als)\, 
    f_{\:\!\rm b}^{}\bigg( \,\frac{x}{y},\,\mufs \bigg)
  \:\: ,
\eeq
form a system of $2n_f\!+\!1$ coupled integro-differential equations.
These equations can be turned into ordinary differential equations by a 
Mellin transformation,
\beq
\label{Mtrf}
  f_{\:\!\rm a\!} \left(N,\mufs \right) 
  \:\:=\:\:
  \int_0^1 \!dx\; x^{\,N-1}\, f_{\:\!\rm a\!} \left(x,\mufs \right)
  \:\: ,
\eeq
and decomposed into $2\:\!\nf\!-\!1$ scalar (non-singlet) equations for the
combinations
\beq
\label{qns}
   q^{\,\pm}_{\rm ik} \;=\; q_{\rm i}^{} \pm \bar{q}_{\rm i}^{} 
                      - ( q_{\rm k}^{} \pm \bar{q}_{\rm k}^{} )
\; , \quad
   q_{\rm v}^{} \;=\; \sum_{\rm i=1}^{\nf} ( q_{\rm i}^{} - \bar{q}_{\rm i}^{} )
\eeq
of quark distributions and the $2 \!\times\! 2$ flavour-singlet 
quark-gluon system
\beq
\label{Sevol}
  \frac{d}{d \ln \mufs}\,
  \left( \begin{array}{c} \!\! q_{\rm s}^{} \!\! \\ \! g\! \end{array} \right)
  \; = \; \left( \begin{array}{cc} P_{\rm qq} & P_{\rm qg} \\
  P_{\rm gq} & P_{\rm gg} \end{array} \right) \otimes
  \left( \begin{array}{c} \!\! q_{\rm s}^{} \!\! \\ \! g\! \end{array} \right)
  \:\: , \quad
  q_s^{} \;=\; \sum_{\rm i=1}^{\nf} ( q_{\rm i}^{} + \bar{q}_{\rm i}^{} )
\eeq
by using the general properties of QCD such as 
$P_{\rm gq_i} = P_{\rm g\bar{q}_i} = P_{\rm gq}$. 
Note that $P_{\rm qg} = 2\:\! \nf P_{\rm q_ig}$.

The splitting functions in Eqs.~(\ref{evol}) admit an expansion in powers of 
$\als$ which we write as
\beq
\label{Pexp}
  P_{\,\rm ij}(x,\als) \; = \; 
  \sum_{n=0} a_{\rm s}^{\,n+1} P^{\,(n)}_{\rm ij}(x)
\quad \mbox{with} \quad
  a_{\rm s} \;=\; \als(\mufs) / (4 \:\!\pi)
\:\: ,
\eeq
i.e., we identify (without loss of information) the mass-factorization and 
the coupling-constant renormalization scales.
The difference between the splitting functions $P_{\rm ns}^{\,+}$ and 
$P_{\rm ns}^{\,-}$ for the first two non-singlet combinations in 
Eq.~(\ref{Sevol}) and the pure-singlet quark-quark splitting function
\beq
\label{Pps}
   P_{\rm ps} \;=\; P_{\rm qq} - P_{\rm ns}^{\,+} 
\eeq
starts at the second order in $\als$, the remaining difference
\beq
\label{PnsS}
   P_{\rm ns}^{\,\rm s} \;=\; P_{\rm ns}^{\,\rm v} - P_{\rm ns}^{\,-}
\eeq
at the third order in $\als$. To order $\as(4)$ the latter quantity is 
proportional to the cubic group invariant $d^{abc\,}d_{abc}/n_c$, while
the other splitting functions can be expressed in terms of $\cf = 4/3$ and
$\ca = n_c = 3$ in QCD and quartic group invariants; the latter do not
occur with the powers of $\nf$ that are considered in this article.
The even-$N$ or odd-$N$ moments of the splitting functions are related to
the anomalous dimensions $\gamma\:\!(N)$ of twist-2 spin-$N$ operators in 
the light-cone operator product expansion (OPE), see, e.g., Refs.\ 
\cite{BurasRev,ReyaRev}; we use the standard convention 
$\,\gamma^{\,(n)}(N) \,=\, - P^{\,(n)}(N)$.

Our calculation of the four-loop splitting functions proceeds along the
lines of Refs.~\cite{Mom3loop1,Mom3loop2,Mom3loop3}.
The partonic DIS structure functions are mapped by the optical theorem to 
forward amplitudes
\beq
\label{ampl}
 \mbox{probe}\,(q)+\mbox{parton}\,(p) \:\longrightarrow\:
 \mbox{probe}\,(q)+\mbox{parton}\,(p) 
\eeq
with $p^2 = 0$ and $q^2 = -Q^2 < 0$. Via a dispersion relation their 
coefficients of $(2p\cdot q/Q^2)^N$ then provide, depending on the structure 
function under consideration, the even-$N$ or odd-$N$ moments of the 
unfactorized partonic structure functions.
These quantities are calculated in dimensional regularization with 
$D = 4 - 2\:\!\ep$, and the $n$-loop splitting functions can be extracted 
from the coefficients of $\ep^{-1} \alpha_{\rm s}^{\,n}$.
For the even-$N$ determination of the splitting functions $P_{\rm ns}^{\,+}$
and $P_{\,\rm ik}$ in Eq.~(\ref{Sevol}) we use the photon and the Higgs
boson in the heavy-top limit as the probes. The splitting functions 
$P_{\rm ns}^{\,-}$ and $P_{\rm ns}^{\,\rm v}$ are determined from the 
odd-$N$ vector$\,$--$\,$axial-vector interference structure function~$F_3$.

The projection on the $N^{\,\rm th}$ power in the parton momentum $p$ leads 
to self-energy integrals that can be solved by the {\sc Forcer} program. 
The complexity of these integrals increases by four if $N$ is increased by 
two. Together with the steep increase of the number of integrals with $N$, 
see the discussion of the harmonic projection in Ref.~\cite{MINCER2}, this
limits the number of moments that can be calculated. So far high
values of $N$ cannot be reached for the top-level 4-loop diagram topologies.

\begin{figure}[bht]
\label{diags1}
\hspace*{3cm}
\includegraphics[bb = 130 110 310 710, scale = 0.5, angle = 270, clip]{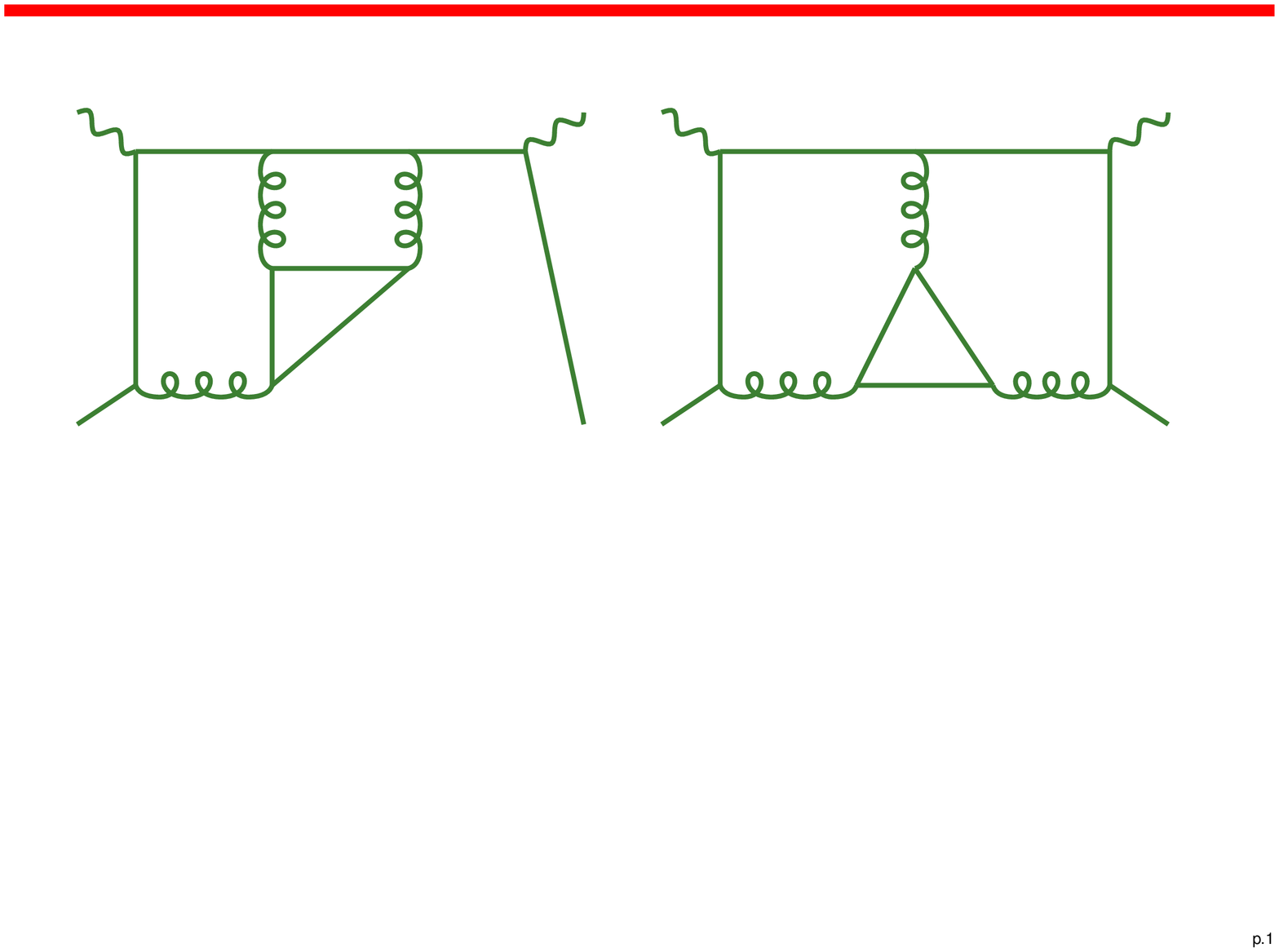}
\vspace{1mm}
\caption{\small 
The three-loop gauge-boson$\,$--$\,$quark forward scattering diagrams with 
{\sc Mincer} topology BE that contribute to $\ca \cf \nf$ part of the
three-loop splitting functions for the quark$\,\pm\,$antiquark flavour 
differences in Eq.~(\ref{qns}).
The same diagrams, but with a one-loop insertion in one of the gluon lines, 
form the hardest part of the corresponding calculation of the four-loop 
$\ca \cf \nfs$ contribution.}
\vspace*{-1mm}
\end{figure}

The raw diagram databases provided by QGRAF \cite{QGRAF} are heavily 
manipulated by (T)FORM \cite{FORM3,TFORM,FORM4} programs to provide the
best possible starting point for the main integral computations. As discussed
in Ref.~\cite{jvLL2016}, one important step is the identification of 
$\ell$-loop self-energy insertions, which reduces many $n$-loop diagrams to 
fewer $(n-\ell)$-loop diagrams in which one or more propagators have a 
non-integer power. For the large-$\nf$ contributions under consideration in 
the article, genuine four-loop diagrams remain after this step only in the
calculation of the $\ca \nft$ part of $P_{\rm qg}^{\,(3)}$, and these
diagrams have a rather simple topology: in the notation of {\sc Mincer} they
are generalizations of the Y3 and O1 three-loop topologies. The hardest 
diagrams occur in the $\ca \cf \nfs$ and $\nfs\, \dabctnc$ non-singlet cases: 
these are three-loop BE topologies with a one-loop gluon propagator, see 
Fig.~\ref{diags1}; the highest $N$ calculated here for any of these is $N=27$. 
%

As far as they are known from fixed-order calculations \cite{mvvPns,mvvPsg}
and all-order resummations of leading large-$\nf$ terms
\cite{LargeNf1,LargeNf2,LargeNf3},
the even-$N$ or odd-$N$ moments of the splitting functions (i.e.~the
anomalous dimensions) can be expressed in terms of simple denominators, 
$D_a^{\,k} = (N+a)^{-k}$
and harmonic sums \cite{Hsums,BKurth} with argument $N$ which are recursively 
defined by
\beq
\label{Hsum1}
  S_{\pm m}(N) \:\:=\; \sum_{i=1}^{N}\: \frac{(\pm 1)^i}{i^{\, m}}
\eeq
and
\beq
\label{Hsum2}
  S_{\pm m_1^{},\,m_2^{},\,\ldots,\,m_d}(N) \:\:=\; \sum_{i=1}^{N}\:
  \frac{(\pm 1)^{i}}{i^{\, m_1^{}}}\: S_{m_2^{},\,\ldots,\,m_d}(i) 
\:\: .
\eeq
The weight $w$ of the harmonic sums is defined by the sum of the absolute 
values of the indices~$m_d$.  Sums up to $w = 2\:\!n-1$ occur in the 
$n$-loop anomalous dimensions, but no sums with an index~$-1$. 
For terms with $D_a^{\,k}$ and$/$or coefficients that include values 
$\zeta_m$ of the Riemann $\zeta$-function (with $m \geq 3$, $\zeta_2$ does 
not occur in these functions), the maximal weight of the sums is reduced by 
$k + m$.

It is, of course, possible that other structures occur in the $n$-loop 
anomalous dimensions at $n \geq 4$ -- already the three-loop DIS coefficient 
functions include terms where special combinations of sums are multiplied by
low positive powers of $N$ \cite{mvvF2L,mvvF3}. 
However, one may expect this to happen at $n=4$ only in the terms with low 
powers of $\nf$ which receive contributions from generically new diagram 
topologies. 
Disregarding new structures and terms with $\zeta_{m \geq 3}$ which are much 
easier to fix from low-$N$ results, a general ansatz for the $n$-loop 
anomalous dimensions then~is
\beq
\label{ansatz}
  \gamma^{\,(n)}(N) \:\: = \:\: 
  \sum_{w=0}^{2n+1} \, c_{00w}^{}\, S_w(N) \,+\, 
  \sum_{\;a^{\phantom{a}}} \: \sum_{k=1}^{2n+1} \: \sum_{w=0}^{2n+1-k}
  c_{akw}^{}\, D_a^{\,k}\, S_w(N)
\:\: ,
\eeq
where $S_w(N)$ is a shorthand for all harmonic sums with weight $w$ and 
$S_0(N) \equiv 1$. The terms with $c_{00w}^{}$ only occur in the quark-quark
and gluon-gluon splitting functions and are restricted by the known 
large-$N$ structure of these functions \cite{PvsCusp,AlbinoBall,DMS05}. 
In all cases the range of the sums is reduced for large-$\nf$ contributions 
in a manner that can be inferred from the results at $n \leq 3$ and from the
prime-factor decompositions of the denominators of the calculated moments.

Even so, Eq.~(\ref{ansatz}) usually includes far too many coefficients for a 
direct determination from as many calculated moments. These coefficients, 
however, are integer modulo some predictable powers of $1/3$ at $n \leq 2$ 
\cite{mvvPns,mvvPsg} and in Refs.~\cite{LargeNf1,LargeNf2,LargeNf3}.
Hence the systems of equations can by turned into Diophantine systems which 
require far fewer equations than unknowns.
Given the present limitations of the calculation of diagrams with BE topology,
this is still not sufficient for the $\nfs$ contributions to the four-loop
non-singlet splitting functions. However, these functions include additional 
structures that facilitate solving these equations with the calculable moments.

The crucial point for the determination of the $\nfs$ parts of 
$\gamma^{\,(3)\pm}_{\,\rm ns}(N)$, already presented in \cite{avLL2016}, 
is to write its colour-factor decomposition in two ways,
\bea
\label{gns3Nf2}
  \gamma^{\,(3)\pm}_{\,\rm ns}(N) \Big|_{\,\nfs} &\!=\!&
  \cf \nfs\, \left\{ \,\cf \; 2A^{(3)}(N) \hspace*{11mm}
  + (\ca-2\cf) B_\pm^{\,(3)}(N) \right\}
\nn \\[-1.5mm] &=&
  \cf \nfs\, \left\{ \,\cf \left( 2A^{(3)}(N)-2B^{\,(3)}_\pm(N) \right) 
 + \; \ca\, B_\pm^{\,(3)}(N) \right\}
\; .
\eea
$A^{(3)}(N)$ is the large-$n_c$ result; it is the same for the even-$N$ ($+$) 
and odd-$N$ ($-$) cases and should include only non-alternating harmonic sums,
i.e., only positive indices in Eqs.~(\ref{Hsum1}) and (\ref{Hsum2}).
Once $A^{(3)}(N)$ is known, it is possible to determine $B_+^{\,(3)}(N)$ and 
$B_-^{\,(3)}(N)$ from the $\cf$ parts in the second line of Eq.~(\ref{gns3Nf2})
which require only two-loop diagrams with one two-loop or two one-loop 
insertions. The corresponding three-loop coefficient, defined as in 
Eq.~(\ref{gns3Nf2}) but with $n_{\!f}^{\,1}$, reads
\bea
\label{A2}
 A^{(2)}(N) &\!\!=\!\!&
      8/3 \* \, \left(
      - 2 \*\, \Ss(1,3) - 4\*\, \Ss(2,2) - 6\*\, \Ss(3,1) + 6\*\, \S(4)
      + 20/3 \*\, (\Ss(1,2) + \, \Ss(2,1)) - ( 11\, - \eta ) \*\, \S(3)
      \right)
\nn \\[0.5mm] & & \mbox{\hspn}
      + \left( -1331/27 - 256/9 \*\, \eta + 64/9 \*\, \eta^2
         + 8 \*\, \eta^3 + 256/9 \*\, D_1^{\,2} - 16 \*\,\z3 \right) \*\, \S(1)
\nn \\[0.5mm] & & \mbox{\hspn}
      + \left( 1246/27 - 32/9 \*\, \eta + 16/3 \*\, \eta^2- 32/3 \*\,
        D_1^{\,2} \right) \*\, \S(2)
      \, - \,  17/2 + 323/54 \*\, \eta 
\nn \\[0.5mm] & & \mbox{\hspn}
        - 248/27 \*\, \eta^2 + 8/9 \*\, \eta^3
      - 4 \*\, \eta^4 
      + 2686/27 \*\, D_1^{\,2} + 152/9 \,\* D_1^{\,3}
      + (12+8 \*\, \eta) \*\, \z3 
\:\: . \quad
\eea
As below, the argument $N$ of the sums is suppressed for brevity.  
$\eta$ is defined in Eq.~(\ref{EtaNu}) below.

We have computed the even and odd moments up to $N = 22$ for the determination
and validation of $A^{(3)}(N)$, and the even-$N$ or odd-$N$ moments up to 
$N = 42$ for $B_+^{\,(3)}(N)$ and $B_-^{\,(3)}(N)$. 
The Diophantine systems have been solved using the LLL-based program in 
Refs.~\cite{axbAlg,Calc} at $N \leq 18\,$ for $A^{(3)}(N)$ with 55 unknowns
and at $\,N \leq 40\,$ for $B_\pm^{\,(3)}(N)$ with 115 unknowns.

For the determination of the $\nfs$ part of $\gamma^{\,(3)\rm s}_{\,\rm ns}(N)$
only the odd moments at $N \leq 25$ were available; the result at $N = 27$
was obtained afterwards and used as a check. As mentioned below 
Eq.~(\ref{PnsS}), the function $\gamma^{\;\rm s}_{\,\rm ns}(N)$ only starts at 
order $\as(3)$. This `leading order' $\dabctnc$ contribution reads
\bea
\label{gnsS2}
 \gamma_{\rm ns}^{\:(2)\rm s}(N) &\!\!=\!\!&
 16\*\, \nf\*\, \dabctnc \*\:
    \Big\{
    ( \Ss(-2,1) - \Ss(1,-2) ) \* (-4\*\,\eta-8\*\,\eta^2)
    - \S(1) \*\, \S(-2) \* (32\*\,\nu-20\*\,\eta-8\*\,\eta^2)
\nn \\ & & \mbox{\hspn}
    + \S(-2) \*\, (32\*\,\nu-36\*\,\eta-28\*\,\eta^2-8\*\,\eta^3)
    + \S(1) \*\, (-32\*\,\nu+26\*\,\eta+56\*\,\eta^2+46\*\,\eta^3+12\*\,\eta^4)
\nn \\[0.5mm] & & \mbox{\hspn}
    + \S(3) \*\, (-2\*\,\eta-4\*\,\eta^2)
    + 32\*\,\nu-32\*\,\eta-60\*\,\eta^2-92\*\,\eta^3-44\*\,\eta^4-8\*\,\eta^5
    \Big\}
\eea
where the result has been rendered more compact by using the abbreviations 
\beq
\label{EtaNu}
  \eta \;\equiv \; \{ N (N+1) \}^{-1} \;=\; D_0\, D_1
\:\: , \quad
  \nu \;\equiv \; \{ (N-1)(N+2) \}^{-1} \;=\; D_{-1}\, D_2 
\:\: .
\eeq
As the overall leading-order quantity $P_{\rm qq}^{\,(0)}$, the splitting 
function corresponding to Eq.~(\ref{gnsS2}) is the same for the present 
initial-state and the final-state (fragmentation distributions) evolution, 
cf.~Refs.~\cite{CFP80,SV96,mmvPnsT}, and invariant  under the $x$-space 
transformation $f(x) \rightarrow x f(1/x)$. The (combinations of) harmonic
sums in Eq.~(\ref{gnsS2}) are `reciprocity respecting' (RR), i.e., their 
Mellin inverses are invariant under the above transformation. The same holds
for the combinations of denominators in Eq.~(\ref{EtaNu}). 
Except for $S_1^{\,2}$ and $S_1^{\,3}$ -- products of RR sums lead to higher 
weight RR sums -- all reciprocity-respecting sums to weight three occur in
Eq.~(\ref{gnsS2}). The list of RR function to this weight has been given in 
Ref.~\cite{VelizTrv} with a slightly different basis choice at $w=3$.

Like the overall NLO anomalous dimensions $\gamma_{\,\rm ns}^{\:(1)\pm}(N)$, 
the next-to-leading order $\dabctnc$ contribution 
$\gamma_{\,\rm ns}^{\:(3)\rm s}(N)$ is not reciprocity-respecting.
However, and this is the crucial point, its RR-breaking part can be calculated 
from Eq.~(\ref{gnsS2}) according to the conjecture of Ref.~\cite{DMS05}. 
For the $\nfs$ contribution addressed here it is given by 
$\,-\frac{2}{3}\,\nf\, \frac{d}{dN}\, \gamma_{\rm ns}^{\:(2)\rm s}(N)$,
where the differentiation can be carried out, for example, via the asymptotic
expansion of the sums, see also Ref.~\cite{JBsums5}. That leaves an unknown 
reciprocity-respecting generalization of the form (\ref{gnsS2}) with 
additional $w=4$ sums which can be chosen as

\pagebreak

\vspace*{-1cm}
\beq
  S_1^{\,4} \, , \;\; S_1 S_3 \, , \;\;
  S_{3,1} - S_{1,3} \, , \;\; S_{-2}^{\,2}
\eeq
and
\beq
  S_{-4}\, , \;\; S_1^{\,2\,} S_{-2} \, , \;\;
  S_1 (S_{-2,1} - S_{1,-2}) \, , \;\; S_{-3,1} + S_{1,-3} -2 S_{1,-2,1} 
\:\: .
\eeq
Including also $\nu^2$ terms, one arrives at a trial function with 79
coefficients, of which as many as 15 can be eliminated by imposing the 
existence of the first moment and the correct values (zero) for its
$\zeta$-function contributions, and 9 can be assumed to vanish 
(all contributions with $S_1^{\,3}$ and $S_1^{\,4}$).
The~remaining 56 coefficients have then been found using the 12 odd moments 
with $3 \leq N \leq 25$.

The correctness of the solution has been verified by the (non-$\zeta$) value
of the first moment and the result at $N=27$. It is possible, though,
to judge `by inspection' whether a solution returned by the Diophantine
equation solver \cite{axbAlg,Calc} is correct. For example, the above solution
is returned~as

\vspace*{-3mm}
\begin{verbatim} 
  A short solution is b[45]
  = 160 372 816 -185 -494 238 52 -64 620 -616 308 112 0 -196 256 12 0 -30 
  208 -282 160 92 -136 96 64 4 0 16 -32 40 -64 0 0 -8 0 22 -32 2 0 24 -40 
  24 -4 24 -24 8 0 0 16 0 -16 12 4 0 0 0
\end{verbatim}

\vspace*{-3mm}
\noindent
where the numbers, ordered by overall weight and the weight of the sums
(the details are not relevant here), are the remaining coefficients $c_{akw}$
in Eq.~(\ref{ansatz}) times $3/32$. 
The factor 3 ensures that the effective coefficients are integer, the factor 
$1/32$ removes some overall powers of 2 introduced by our choice for the 
expansion parameter $a_{\rm s}$ in Eq.~(\ref{Pexp}).

A pattern such as the one above for the about 30 coefficients of the 
highest-weight functions, with larger and more random coefficients at the
left (low-weight) end, is a hallmark of a correct solution. In fact, correct 
and incorrect solutions were correctly identified by inspection in all present
calculations as well as in the preparation of Ref.~\cite{mvvDP2}.
 
Of the $\nft$ contributions to the singlet splitting functions in 
Eq.~(\ref{Sevol}), only the case of $P_{\,\rm qg}^{\,(3)}$ is critical.
Unlike the other three cases this function is suppressed by only two powers
of $\nf$ relative to the lowest-$\nf$ term, recall the remark below 
Eq.~(\ref{Sevol}), and includes contributions from sums up to weight four
instead of weight three. Hence a considerably larger basis set is required
in Eq.~(\ref{ansatz}). At the same time the fixed-$N$ calculations are 
harder for $P_{\,\rm qg}^{\,(3)}$ than for the other three cases, in 
particular for the $\ca \nft$ contribution, as already indicated on p.~4.
  
Yet, using reasonable assumptions based on the three-loop splitting function,
we managed to find suitable functional forms with 101 unknown coefficients 
for the $\cf \nft$ part (with only positive-index sums but overall 
weight up to six) and 115 unknown coefficients for the $\ca \nft$ part 
(including alternating sums but an overall weight of five), which we were 
able to determine from the even moments $2 \leq N \leq 40$ in the former and 
$2 \leq N \leq 44$ in the latter case. Several higher moments were employed
for the validation of the $\cf \nft$ result and the $\ca \nft$ coefficients 
were checked using $N = 46$. Some of the four-loop and three-loop $\ca \nft$ 
diagrams at $N > 40$ were calculated using an alternative approach 
for generalized Y and  O {\sc Mincer} topologies that avoids the harmonic 
projection \cite{MINCER2}. 
This approach may be reported on later in a more general context.

%
\setcounter{equation}{0}
\section{Results in $N$-space}
\label{sec:Nres}

In this section we present the analytic expressions for the $\nfs$ and 
$\nft$ contributions to the three non-singlet anomalous dimensions and
the $\nft$ parts of their four flavour-singlet counterparts in the
\MSb\ scheme.
As~in Eqs.~(\ref{A2}) and (\ref{gnsS2}) above, all harmonic sums 
(\ref{Hsum1}) and (\ref{Hsum2}) have the argument $N$ which is suppressed 
in the formulae for brevity.

The results for $\gamma^{\,(3)\pm}_{\,\rm ns}$ are presented in terms of 
the decomposition (\ref{gns3Nf2}). The large-$n_c$ part 
\bea
\label{eqn:A}
\lefteqn{A^{(3)}(N) \;=\;
\frct{16}{27}\* \,\bigg\{
       - 12\, \* \Sss(1,3,1)\,
       + 6\, \* \Ss(1,4)\,
       - 12\, \* \Ss(2,3)\,
       - 24\, \* \Ss(3,2)\,
       - 30\, \* \Ss(4,1)\,
       + 36\, \* \S(5)\,
       + 20\, \* \Ss(1,3)\,
}
   \nn \\[0.0mm] & & \mbox{}
       + 40\, \* \Ss(2,2)\,
       + 6\, \* \Ss(3,1)\, \* \Big(10 + \eta\Big)\,
       - 3/2\: \* \S(4)\, \* \Big(53 + 2\, \* \eta\Big)\,
       - 38/3\: \* \Ss(1,2)\,
       - 38/3\: \* \Ss(2,1)\,
   \nn \\[0.0mm] & & \mbox{}
       + 1/3\: \* \S(3)\, \* \Big(287 - 12\, \* \eta\,
              + 18\, \* \eta^{2}\,
              - 36\, \* D_{1}^{2}\Big)\,
       - 1/12\: \* \S(2)\, \* \Big(416\, \* \eta - 12\, \* \eta^{2}\,
              - 144\, \* \eta^{3}\,
   \nn \\[0.0mm] & & \mbox{}
              - 768\, \* D_{1}^{2}\,
              + (1259\,
              + 216\, \* \zeta_3)\Big)\,
       + 1/48\: \* \S(1)\, \* \Big(3392\, \* \eta - 3656\, \* \eta^{2}\,
              + 432\, \* \eta^{3}\,
   \nn \\[0.0mm] & & \mbox{}
              + 720\, \* \eta^{4}\,
              - 3392\, \* D_{1}^{2}\,
              - 576\, \* D_{1}^{3}\,
              - 1728\, \* D_{1}^{4}\,
              + (2119\,
              + 2880\, \* \zeta_3\,
              - 1296\, \* \zeta_4)\Big)\,
   \nn \\[0.0mm] & & \mbox{}
       + 1/96\: \* \Big(944\, \* \eta^{3} - 864\, \* \eta^{5}\,
              - 7088\, \* D_{1}^{3}\,
              - 2736\, \* D_{1}^{4}\,
              - 1728\, \* D_{1}^{5}\,
              + 9\, \* (127\,
              - 264\, \* \zeta_3\,
   \nn \\[0.5mm] & & \mbox{}
              + 216\, \* \zeta_4)\,
              - 24\, \* (1705\,
              + 72\, \* \zeta_3)\, \* D_{1}^{2}\,
              - 2\, \* (2275\,
              - 432\, \* \zeta_3)\, \* \eta^{2}\,
              + (20681\,
              - 2880\, \* \zeta_3\,
   \nn \\[0.5mm] & & \mbox{}
              + 1296\, \* \zeta_4)\, \* \eta\Big)
\bigg\}
\eea
is the same for these two cases, while the contributions with the
$1/n_c$-suppressed `non-planar' colour factor $(\ca - 2\,\cf)$ are 
valid at even $N$ for $B_+^{\,(3)}$ and odd $N$ for $B_-^{\,(3)}$.
These functions read
\bea
\label{eqn:Bp}
\lefteqn{B_+^{\,(3)}(N) \;=\;
\frct{32}{27}\*\, \bigg\{
       - 9\, \* \S(-5)\,
       - 12\, \* \Ss(-4,1)\,
       - 6\, \* \Ss(-3,-2)\,
       - 12\, \* \Sss(-3,1,1)\,
       + 6\, \* \Ss(1,-4)\,
       + 12\, \* \Sss(1,-3,1)\,
}
   \nn \\[0.5mm] & & \mbox{}
       + 12\, \* \Sss(1,-2,-2)\,
       + 24\, \* \Ssss(1,-2,1,1)\,
       - 6\, \* \Sss(1,3,1)\,
       + 24\, \* \Ss(1,4)\,
       + 6\, \* \Ss(2,-3)\,
       + 12\, \* \Sss(2,-2,1)\,
       + 9\, \* \Ss(2,3)\,
   \nn \\[0.0mm] & & \mbox{}
       + 6\, \* \Ss(3,-2)\,
       - 3\, \* \Ss(3,2)\,
       - 6\, \* \Ss(4,1)\,
       + 9\, \* \S(5)\,
       + \S(-4)\, \* \Big(20 - 3\, \* \eta\Big)\,
       + 2\, \* \Ss(-3,1)\, \* \Big(10 - 3\, \* \eta\Big)\,
   \nn \\[0.5mm] & & \mbox{}
       - 6\, \* \Ss(-2,-2)\, \* \eta\,
       - 12\, \* \Sss(-2,1,1)\, \* \eta\,
       - 20\, \* \Ss(1,-3)\,
       - 40\, \* \Sss(1,-2,1)\,
       - 30\, \* \Ss(1,3)\,
       - 20\, \* \Ss(2,-2)\,
   \nn \\[0.5mm] & & \mbox{}
       + \Ss(3,1)\, \* \Big(10 + 3\, \* \eta\Big)\,
       - 1/2\: \* \S(4)\, \* \Big(73 + 24\, \* \eta\Big)\,
       - 1/3\: \* \S(-3)\, \* \Big(19 - 30\, \* \eta\,
              + 9\, \* \eta^{2}\,
              - 18\, \* D_{1}^{2}\Big)\,
   \nn \\[0.0mm] & & \mbox{}
       + 2\, \* \Ss(-2,1)\, \* \Big(10\, \* \eta - 3\, \* \eta^{2}\,
              + 6\, \* D_{1}^{2}\Big)\,
       + 38/3\: \* \Ss(1,-2)\,
       + 1/12\: \* \S(3)\, \* \Big(619 + 180\, \* \eta\,
   \nn \\[0.5mm] & & \mbox{}
              - 54\, \* \eta^{2}\,
              + 108\, \* D_{1}^{2}\Big)\,
       + 1/3\: \* \S(-2)\, \* \Big(8\, \* \eta + 39\, \* \eta^{2}\,
              - 96\, \* D_{1}^{2}\Big)\,
       + 6\, \* \Ss(1,1)\, \* \Big(2\, \* \eta^{2} + \eta^{3}\Big)\,
   \nn \\[0.0mm] & & \mbox{}
       + 1/48\: \* \S(2)\, \* \Big(144\, \* \eta^{2} + 72\, \* \eta^{3}\,
              - (1585\,
              + 864\, \* \zeta_3)\Big)\,
       + 1/96\: \* \S(1)\, \* \Big(1584\, \* \eta - 3672\, \* \eta^{2}\,
   \nn \\[0.5mm] & & \mbox{}
              + 720\, \* \eta^{3}\,
              + 864\, \* \eta^{4}\,
              - 1728\, \* D_{1}^{2}\,
              - 1728\, \* D_{1}^{3}\,
              - 2592\, \* D_{1}^{4}\,
              + (923\,
              + 5760\, \* \zeta_3\,
   \nn \\[0.5mm] & & \mbox{}
              - 2592\, \* \zeta_4)\Big)\,
       - 1/192\: \* \Big(1392\, \* \eta^{3} - 1584\, \* \eta^{4}\,
              + 3168\, \* D_{1}^{4}\,
              - 3\, \* (193\,
              - 1584\, \* \zeta_3\,
   \nn \\[0.5mm] & & \mbox{}
              + 1296\, \* \zeta_4)\,
              + 2\, \* (2447\,
              - 864\, \* \zeta_3)\, \* \eta^{2}\,
              + 4\, \* (7561\,
              + 864\, \* \zeta_3)\, \* D_{1}^{2}\,
              - (15077\,
              - 5760\, \* \zeta_3\,
   \nn \\[0.5mm] & & \mbox{}
              + 2592\, \* \zeta_4)\, \* \eta\Big)
       \bigg\}
\eea
and
\bea
\label{eqn:Bm}
\lefteqn{B_-^{\,(3)}(N) \;=\;
\frct{32}{27}\*\, \bigg\{
       - 9\, \* \S(-5)\,
       - 12\, \* \Ss(-4,1)\,
       - 6\, \* \Ss(-3,-2)\,
       - 12\, \* \Sss(-3,1,1)\,
       + 6\, \* \Ss(1,-4)\,
       + 12\, \* \Sss(1,-3,1)\,
}
   \nn \\[0.5mm] & & \mbox{}
       + 12\, \* \Sss(1,-2,-2)\,
       + 24\, \* \Ssss(1,-2,1,1)\,
       - 6\, \* \Sss(1,3,1)\,
       + 24\, \* \Ss(1,4)\,
       + 6\, \* \Ss(2,-3)\,
       + 12\, \* \Sss(2,-2,1)\,
       + 9\, \* \Ss(2,3)\,
   \nn \\[0.5mm] & & \mbox{}
       + 6\, \* \Ss(3,-2)\,
       - 3\, \* \Ss(3,2)\,
       - 6\, \* \Ss(4,1)\,
       + 9\, \* \S(5)\,
       + \S(-4)\, \* \Big(20 - 3\, \* \eta\Big)\,
       + 2\, \* \Ss(-3,1)\, \* \Big(10 - 3\, \* \eta\Big)\,
   \nn \\[0.5mm] & & \mbox{}
       - 6\, \* \Ss(-2,-2)\, \* \eta\,
       - 12\, \* \Sss(-2,1,1)\, \* \eta\,
       - 20\, \* \Ss(1,-3)\,
       - 40\, \* \Sss(1,-2,1)\,
       - 30\, \* \Ss(1,3)\,
       - 20\, \* \Ss(2,-2)\,
   \nn \\[0.5mm] & & \mbox{}
       + \Ss(3,1)\, \* \Big(10 + 3\, \* \eta\Big)\,
       - 1/2\: \* \S(4)\, \* \Big(73 + 24\, \* \eta\Big)\,
       - 1/3\: \* \S(-3)\, \* \Big(19 - 30\, \* \eta\,
              + 9\, \* \eta^{2}\,
              - 18\, \* D_{1}^{2}\Big)\,
   \nn \\[0.0mm] & & \mbox{}
       + 2\, \* \Ss(-2,1)\, \* \Big(10\, \* \eta - 3\, \* \eta^{2}\,
              + 6\, \* D_{1}^{2}\Big)\,
       + 38/3\: \* \Ss(1,-2)\,
       + 1/12\: \* \S(3)\, \* \Big(619 + 180\, \* \eta\,
              - 54\, \* \eta^{2}\,
   \nn \\[0.0mm] & & \mbox{}
              + 108\, \* D_{1}^{2}\Big)\,
       + 1/3\: \* \S(-2)\, \* \Big(8\, \* \eta + 3\, \* \eta^{2}\,
              - 18\, \* \eta^{3}\,
              - 96\, \* D_{1}^{2}\Big)\,
       - 6\, \* \Ss(1,1)\, \* \Big(2\, \* \eta^{2} + \eta^{3}\Big)\,
   \nn \\[0.0mm] & & \mbox{}
       + 1/48\: \* \S(2)\, \* \Big(144\, \* \eta^{2} + 72\, \* \eta^{3}\,
              - (1585\,
              + 864\, \* \zeta_3)\Big)\,
       - 1/96\: \* \S(1)\, \* \Big(432\, \* \eta - 1032\, \* \eta^{2}\,
   \nn \\[0.0mm] & & \mbox{}
              + 240\, \* \eta^{3}\,
              + 288\, \* \eta^{4}\,
              - 576\, \* D_{1}^{2}\,
              - 576\, \* D_{1}^{3}\,
              - 864\, \* D_{1}^{4}\,
              - (923\,
              + 5760\, \* \zeta_3\,
              - 2592\, \* \zeta_4)\Big)\,
   \nn \\[0.5mm] & & \mbox{}
       + 1/192\: \* \Big(7280\, \* \eta^{3} - 336\, \* \eta^{4}\,
              - 1728\, \* \eta^{5}\,
              - 11136\, \* D_{1}^{3}\,
              - 18144\, \* D_{1}^{4}\,
              + 4608\, \* D_{1}^{5}\,
   \nn \\[0.5mm] & & \mbox{}
              + 3\, \* (193\,
              - 1584\, \* \zeta_3\,
              + 1296\, \* \zeta_4)\,
              - 18\, \* (583\,
              - 96\, \* \zeta_3)\, \* \eta^{2}\,
              - 4\, \* (10489\,
              + 864\, \* \zeta_3)\, \* D_{1}^{2}\,
   \nn \\[0.5mm] & & \mbox{}
              + (25541\,
              - 5760\, \* \zeta_3\,
              + 2592\, \* \zeta_4)\, \* \eta\Big)\,
\bigg\}
\:\: .
\eea
As for the complete corresponding three-loop quantities in Ref.~\cite{mvvPsg},
the difference between the odd-$N$ result (\ref{eqn:Bm}) and the even-$N$
result (\ref{eqn:Bp}) is much simpler than those expressions and given by 
\bea
\label{eqn:dB}
\lefteqn{\delta B^{\,(3)}(N) \;=\;
   \frct{32}{27}\*\, \bigg\{
       - 6\, \* \S(-2)\, \* \Big(2\, \* \eta^{2} + \eta^{3}\Big)\,
       - 12\, \* \Ss(1,1)\, \* \Big(2\, \* \eta^{2} + \eta^{3}\Big)\,
       - \S(1)\, \* \Big(21\, \* \eta - 49\, \* \eta^{2}\,
}
   \nn \\[0.0mm] & & \mbox{}
              + 10\, \* \eta^{3}\,
              + 12\, \* \eta^{4}\,
              - 24\, \* D_{1}^{2}\,
              - 24\, \* D_{1}^{3}\,
              - 36\, \* D_{1}^{4}\Big)\,
       + 1/6\: \* \Big(327\, \* \eta - 175\, \* \eta^{2}\,
              + 271\, \* \eta^{3}\,
   \nn \\[0.0mm] & & \mbox{}
              - 60\, \* \eta^{4}\,
              - 54\, \* \eta^{5}\,
              - 366\, \* D_{1}^{2}\,
              - 348\, \* D_{1}^{3}\,
              - 468\, \* D_{1}^{4}\,
              + 144\, \* D_{1}^{5}\Big)
\bigg\}
\:\: .
\eea
Finally the additional $\nfs \ar(4)$ contribution to the evolution of the 
valence distribution, see Eq.~(\ref{PnsS}),~is
\bea
\label{eqn:S}
\lefteqn{ \gamma_{\,\rm ns}^{\,(3)\rm s}\big|_{\nfs \dabctnc}(N) \;=\;
\frct{64}{3}\*\, \bigg\{
        2\, \* \Big[\S(-4)\,
              + 2\, \* \Ss(-3,1)\,
              + 2\, \* \Ss(1,-3)\,
              - 4\, \* \Sss(1,-2,1)\,
              - \Ss(1,3)\Big]\, \* \Big(8\, \* \nu - 5\, \* \eta\,
}
   \nn \\[0.0mm] & & \mbox{}
              - 2\, \* \eta^{2}\Big)\,
       - 8\, \* \Big[2\, \* \Ss(-2,-2)\,
              + 4\, \* \Sss(-2,1,1)\,
              - \Ss(-2,2) \Big]\, \* \Big(2\, \* \nu - \eta\Big)\,
       - 4\, \* \Big[4\, \* \Sss(1,1,-2)\,
              - \Ss(2,-2)\,
   \nn \\[0.0mm] & & \mbox{}
              + \Ss(3,1)\Big]\, \* \Big(4\, \* \nu - 3\, \* \eta\,
              - 2\, \* \eta^{2}\Big)\,
       + 2\, \* \S(4)\, \* \Big(16\, \* \nu - 11\, \* \eta\,
              - 6\, \* \eta^{2}\Big)\,
       - 2/3\: \* \S(-3)\, \* \Big(128\, \* \nu - 87\, \* \eta\,
   \nn \\[0.0mm] & & \mbox{}
              - 21\, \* \eta^{2}\,
              + 6\, \* \eta^{3}\,
              - 6\, \* D_{1}^{2}\,
              + 24\, \* D_{1}^{3}\,
              + 16\, \* D_{2}^{2}\Big)\,
       + 4/3\: \* \Ss(-2,1)\, \* \Big(88\, \* \nu - 57\, \* \eta\,
              - 21\, \* \eta^{2}\,
              - 6\, \* \eta^{3}\,
   \nn \\[0.0mm] & & \mbox{}
              - 12\, \* D_{1}^{2}\,
              + 8\, \* D_{2}^{2}\Big)\,
       + 8/3\: \* \Ss(1,-2)\, \* \Big(44\, \* \nu - 42\, \* \eta\,
              - 21\, \* \eta^{2}\,
              + 3\, \* D_{1}^{2}\,
              + 12\, \* D_{1}^{3}\,
              + 4\, \* D_{2}^{2}\Big)\,
   \nn \\[0.0mm] & & \mbox{}
       + \S(3)\, \* \Big(16\, \* \nu - 9\, \* \eta\,
              - 7\, \* \eta^{2}\,
              - 6\, \* \eta^{3}\,
              - 6\, \* D_{1}^{2}\,
              - 8\, \* D_{1}^{3}\Big)\,
       - 1/3\: \* \S(-2)\, \* \Big(304\, \* \nu - 273\, \* \eta\,
   \nn \\[0.0mm] & & \mbox{}
              - 312\, \* \eta^{2}\,
              - 84\, \* \eta^{3}\,
              - 84\, \* D_{1}^{2}\,
              + 24\, \* D_{1}^{3}\,
              - 72\, \* D_{1}^{4}\,
              + 32\, \* D_{2}^{2}\Big)\,
       - \Big[4\, \* \Ss(1,1) 
                 - \S(2)\Big] \* \Big(16\, \* \nu 
   \nn \\[0.0mm] & & \mbox{}
              - 13\, \* \eta\,
              - 28\, \* \eta^{2}\,
              - 23\, \* \eta^{3}\,
              - 6\, \* \eta^{4}\Big)\,
       + 1/6\: \* \S(1)\, \* \Big(608\, \* \nu - 855\, \* \eta\,
              - 984\, \* \eta^{2}\,
              - 972\, \* \eta^{3}\,
   \nn \\[0.0mm] & & \mbox{}
              - 144\, \* \eta^{4}\,
              + 24\, \* \eta^{5}\,
              + 300\, \* D_{1}^{2}\,
              + 456\, \* D_{1}^{3}\,
              + 36\, \* D_{1}^{4}\,
              + 288\, \* D_{1}^{5}\,
              + 64\, \* D_{2}^{2}\Big)\,
   \nn \\[0.0mm] & & \mbox{}
       - 2/3\: \* \Big(104\, \* \nu + 96\, \* \eta\,
              - 261\, \* \eta^{2}\,
              - 252\, \* \eta^{3}\,
              - 54\, \* \eta^{4}\,
              + 36\, \* \eta^{5}\,
              + 12\, \* \eta^{6}\,
              - 216\, \* D_{1}^{2}\,
   \nn \\[0.0mm] & & \mbox{}
              - 168\, \* D_{1}^{3}\,
              - 162\, \* D_{1}^{4}\,
              + 24\, \* D_{1}^{5}\,
              - 60\, \* D_{1}^{6}\,
              + 16\, \* D_{2}^{2}\Big)
\bigg\}
\:\: .
\eea
The leading large-$\nf$ contribution is the same for the three types of
non-singlet quark distributions in Eq.~(\ref{qns}). It has been obtained 
to all orders in $\als$ in Ref.~\cite{LargeNf1}. 
Our results agree with the corresponding fourth-order coefficient which 
in our notation reads, for even and odd $N$,
\bea
\label{gns3nf3}
\lefteqn{\gqqns{3}\big|_{\nft}(N) \;=\,
        \frct{16}{81}\*\, \cf \,\* \Big\{
           6\,\* \S(4)\,
         - 10\, \* \S(3)\,
         - 2\,\* \S(2)\,
         - \S(1)\, \* (2 - 12\, \* \zeta_3)\,
         + 131/16 
}
\hspace*{8mm}
   \nn \\[0.0mm] & & \mbox{\hspp\hspp} \vphantom{\Big(}
   - 9\,\* \z3
         - \eta \*\, ( 20 + 6 \*\, \z3 )
         + 15\*\, \eta^2
         - \eta^3 
         - 3 \*\, \eta^4
         + 24\, \* D_{1}^{2}
         + 6\, \* D_{1}^{4}
         \Big\}
\:\: .
\eea

The new functions (\ref{eqn:A}) -- (\ref{eqn:S}) are illustrated in Fig.~2.
The results at non-integer values of $N$ have been calculated by a numerical
Mellin transformation of the $x$-space expressions in the next section; 
for~the analytic continuation in $N$ of the harmonic sums to weight five 
see also 
Ref.~\cite{JBsums5}.

Up to terms suppressed by two powers of $1/N$, also the large-$N$ behaviour 
of the three non-singlet anomalous dimension is the same with
\beq
\label{ntoinf}
  \gamma_{\,\rm ns}^{\:(n-1)}(N) \; = \:\; 
    A_n \, (\ln N +\gamma_{\:\!e}) - B_n 
  + C_n \;\frac{\ln N + \gamma_{\:\!e}}{N} 
  - D_n + {\cal O} \left( {N^{\,-2} \ln^{\,\ell\!} N} \right)
\eeq
where $\gamma_{\:\!e}$ is the Euler-Mascheroni constant. 
The coefficients $A_n$ are relevant beyond the evolution of the parton 
distributions, since they are identical to the $n$-loop cusp anomalous 
dimensions \cite{PvsCusp}. 
The result at three loops can be found in Eq.~(3.11) of Ref.~\cite{mvvPns}, 
its $\nf$ part was derived before in Refs.~\cite{mvvNSnf,BergerA3nf}.
Our new results (\ref{eqn:A}) and (\ref{eqn:Bp}) specify the $\nfs$
coefficient of $A_4$. Together with the long-known $\nft$ result 
\cite{LargeNf1,BB95nf} given by the large-$N$ limit of Eq.~(\ref{gns3nf3})
we obtain
\bea
\label{A4nf23}
           A_4 \Big|_{n_{\!f}^{\;a\,>1}}
  &\!\!=\!& 
           \cf \* \ca \* \nfs \* \left(
             \frac{923}{81}
           - \frac{608}{81}\,\* \z2
           + \frac{2240}{27}\,\* \z3
           - \frac{112}{3}\,\* \z4
           \right)
\nn \\[0.5mm] && \mbox{\hspn}
       +\: \cfs \* \nfs \* \left(
             \frac{2392}{81}
           - \frac{640}{9}\,\* \z3
           + 32\,\* \z4
           \right)
         - \cf \* \nft \* \left(
             \frac{32}{81}
           - \frac{64}{27}\,\* \z3
           \right)
\:\! .
\eea
The large-$n_c$ limit of this result has also been derived in Ref.~%
\cite{HSSS16}, and the $\cfs \nfs$ part in Refs.\ \cite{GHKM15,GrozinLL}. 
Hence all $\nfs$ contributions in Eq.~(\ref{A4nf23}) are covered by 
two independent determinations. Since~this result involves several 
coefficients in (\ref{eqn:A}) and (\ref{eqn:Bp}), this agreement can also
be viewed as another verification of our determination of the all-$N$ $\nfs$
expressions for $\gamma_{\,\rm ns}^{\:(3)\pm}$.
The $\nfs$ part of the coefficient $C_4$ in Eq.~(\ref{ntoinf}) 
is found to be
\beq
\label{C4nf2}
   C_4 \Big|_\nfs \;=\; \frac{1216}{81}\; \cfs \nfs
   \;=\; \Big[ (A_2)^2 + A_1 A_3 \Big]_\nfs
\eeq
as conjectured in Ref.~\cite{DMS05} -- 
the first verification of this conjecture by a fourth-order calculation.

\pagebreak

\begin{figure}[tb]
\vspace{-2mm}
\centerline{\epsfig{file=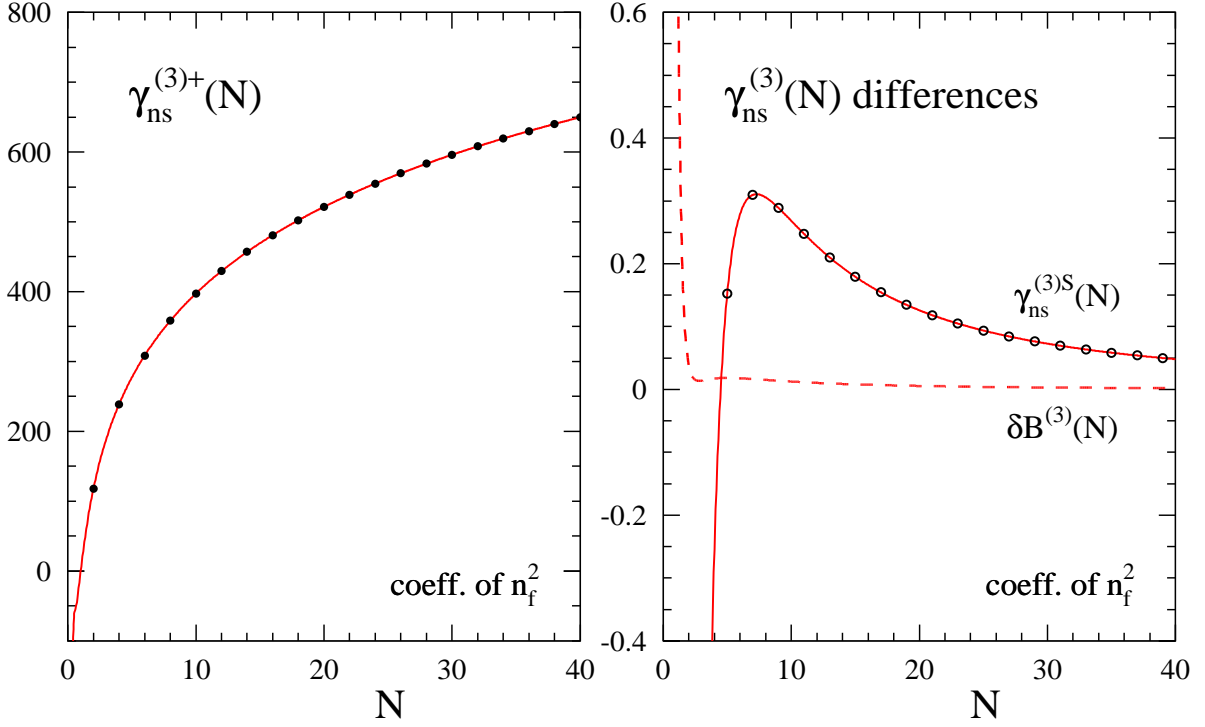,width=16.0cm,angle=0}}
\vspace{-2mm}
\caption{ \label{fig:gnsnf2} \small
 The $\nfs$ parts of the anomalous dimensions $\gamma_{\,\rm ns}^{\:(3)+}(N)$
 (left) and $\gamma_{\,\rm ns}^{\:(3)\rm s}(N) = 
 \gamma_{\,\rm ns}^{\:(3)\rm v}(N) - \gamma_{\,\rm ns}^{\:(3)-}(N)$ (right).
 Their even-$N$ (left) and odd-$N$ (odd) moments computed using {\sc Forcer}
 \cite{tuLL2016,FORCER} are shown together with the numerical all-$N$ curves.
 Also shown on the right, where we focus on $\gamma_{\,\rm ns}^{\:(3)\rm s}$ at 
 $N > 4$, is the difference $\delta B^{\,(3)}(N) = \gamma_{\,\rm ns}^{\:(3)-}(N) 
 - \gamma_{\,\rm ns}^{\:(3)+}(N)$. Note the normalization of our expansion 
 parameter~$a_{\rm s}$ in Eq.~(\ref{Pexp}). }
\vspace{-1mm}
\end{figure}

We now turn to the leading large-$\nf$ anomalous dimensions for the even-$N$
flavour-singlet evolution (\ref{Sevol}), starting with the pure singlet
contribution (\ref{Pps}):
\bea
\label{gpsnf3}
\lefteqn{\gqqps{3}\big|_{\nft}(N) \;=\,
          \colourcolour{\cf} \, \*  \Big\{\,
          - 64/27\: \* \Sss(1,1,1)\, \* \Big(3\, \* D_{0}\,
              - 6\, \* D_{0}^{2}\,
              - 3\, \* D_{1}\,
              - 6\, \* D_{1}^{2}\,
              - 4\, \* D_{2}\,
              + 4\, \* D_{-1}\Big)\,
}
   \nn \\[0mm] & & \mbox{} \vphantom{\Big(}
          + 64/27\: \* \Ss(1,1)\, \* \Big(11\, \* D_{0}\,
              - 13\, \* D_{0}^{2}\,
              + 6\, \* D_{0}^{3}\,
              - 17\, \* D_{1}\,
              - 4\, \* D_{1}^{2}\,
              + 12\, \* D_{1}^{3}\,
              + 2\, \* D_{2}\,
              + 8\, \* D_{2}^{2}\,
   \nn \\[0mm] & & \mbox{} \vphantom{\Big(}
              + 4\, \* D_{-1}\Big)\,
          - 32/81\: \* \S(1)\, \* \Big(94\, \* D_{0}\,
              - 98\, \* D_{0}^{2}\,
              + 87\, \* D_{0}^{3}\,
              - 18\, \* D_{0}^{4}\,
              - 226\, \* D_{1}\,
              + 100\, \* D_{1}^{2}\,
   \nn \\[0mm] & & \mbox{} \vphantom{\Big(}
              + 111\, \* D_{1}^{3}\,
              - 90\, \* D_{1}^{4}\,
              + 128\, \* D_{2}\,
              + 88\, \* D_{2}^{2}\,
              - 48\, \* D_{2}^{3}\,
              + 4\, \* D_{-1}\Big)\,
          + 16/81\: \* \Big(146\, \* D_{0}^{3}\,
   \nn \\[0mm] & & \mbox{} \vphantom{\Big(}
              - 87\, \* D_{0}^{4}\,
              + 18\, \* D_{0}^{5}\,
              - 54\, \* D_{1}^{3}\,
              - 309\, \* D_{1}^{4}\,
              + 198\, \* D_{1}^{5}\,
              + 72\, \* D_{2}^{2}\,
              - 176\, \* D_{2}^{3}\,
              + 96\, \* D_{2}^{4}\,
   \nn \\[0mm] & & \mbox{} \vphantom{\Big(}
              - 4\, \* (1 - 18\, \* \z3)\, \* D_{-1}\,
              + 2\, \* (26 + 27\, \* \z3)\, \* D_{0}\,
              - 2\, \* (59 + 54\, \* \z3)\, \* D_{0}^{2}\,
   \nn \\[0mm] & & \mbox{} \vphantom{\Big(}
              + 4\, \* (91 - 18\, \* \z3)\, \* D_{2}\,
              - 2\, \* (206 + 27\, \* \z3)\, \* D_{1}\,
              + 2\, \* (215 - 54\, \* \z3)\, \* D_{1}^{2}\Big)
          \Big\}
\:\: .
\eea
As expected from the lower orders, the highest-weight sums in the four-loop
off-diagonal contributions are proportional to the leading-order structures
\beq
\label{pqgpgq}
  p_{\rm qg} \;=\; D_0 - 2\*\, D_1 + 2\*\, D_2 
\quad \mbox{and} \quad
  p_{\rm gq} \;=\; 2\*\, D_{-1} - 2\*\, D_0 + D_1
\:\: .
\eeq
Using these abbreviations, the fourth-order leading-$\nf$ parts 
of the gluon-quark and quark-gluon anomalous dimensions are given by
 
\pagebreak

\vspace*{-1cm}
\bea
\label{gqgnf3}
\lefteqn{\gqg{3}\big|_{\nft}(N) \;=\,
          \colourcolour{\cf} \, \*  \Big\{\,
          32/27\: \* \Big[3\, \* \S(4) - \Ssss(1,1,1,1)\Big]\, \* \npqg
          - 32/81\: \* \Sss(1,1,1)\, \* \Big(71\, \* D_{0}\,
              - 30\, \* D_{0}^{2}\,
              + 18\, \* D_{0}^{3}\,
}
   \nn \\[0mm] & & \mbox{} \vphantom{\Big(}
              - 115\, \* D_{1}\,
              - 36\, \* D_{1}^{3}\,
              + 42\, \* D_{2}\,
              + 24\, \* D_{2}^{2}\,
              - 8\, \* D_{-1}\Big)\,
          + 32/81\: \* \Big[\Ss(1,2)+\Ss(2,1)\Big]\, \* \Big(81\, \* D_{0}\,
   \nn \\[0mm] & & \mbox{} \vphantom{\Big(}
              - 27\, \* D_{0}^{2}\,
              + 18\, \* D_{0}^{3}\,
              - 135\, \* D_{1}\,
              - 36\, \* D_{1}^{3}\,
              + 62\, \* D_{2}\,
              + 24\, \* D_{2}^{2}\,
              - 8\, \* D_{-1}\Big)\,
   \nn \\[0mm] & & \mbox{} \vphantom{\Big(}
          + 32/81\: \* \S(3)\, \* \Big(71\, \* D_{0}\,
              - 27\, \* D_{0}^{2}\,
              + 18\, \* D_{0}^{3}\,
              - 109\, \* D_{1}\,
              - 36\, \* D_{1}^{3}\,
              + 36\, \* D_{2}\,
              + 24\, \* D_{2}^{2}\,
              - 8\, \* D_{-1}\Big)\,
   \nn \\[0mm] & & \mbox{} \vphantom{\Big(}
          - 16/243\: \* \Ss(1,1)\, \* \Big(416\, \* D_{0}\,
              - 102\, \* D_{0}^{2}\,
              - 72\, \* D_{0}^{3}\,
              - 1633\, \* D_{1}\,
              + 90\, \* D_{1}^{2}\,
              - 288\, \* D_{1}^{3}\,
              - 216\, \* D_{1}^{4}\,
   \nn \\[0mm] & & \mbox{} \vphantom{\Big(}
              + 1174\, \* D_{2}\,
              + 648\, \* D_{2}^{2}\,
              + 288\, \* D_{2}^{3}\,
              + 72\, \* D_{-1}\Big)\,
          - 32/243\: \* \S(2)\, \* \Big(976\, \* D_{0}\,
              - 891\, \* D_{0}^{2}\,
   \nn \\[0mm] & & \mbox{} \vphantom{\Big(}
              + 360\, \* D_{0}^{3}\,
              - 216\, \* D_{0}^{4}\,
              + 88\, \* D_{1}\,
              - 459\, \* D_{1}^{2}\,
              - 72\, \* D_{1}^{3}\,
              + 540\, \* D_{1}^{4}\,
              - 1101\, \* D_{2}\,
              - 852\, \* D_{2}^{2}\,
   \nn \\[0mm] & & \mbox{} \vphantom{\Big(}
              - 432\, \* D_{2}^{3}\,
              + 68\, \* D_{-1}\Big)\,
          - 16/729\: \* \S(1)\, \* \Big(8634\, \* D_{0}^{2}\,
              - 6822\, \* D_{0}^{3}\,
              + 2430\, \* D_{0}^{4}\,
              - 1620\, \* D_{0}^{5}\,
   \nn \\[0mm] & & \mbox{} \vphantom{\Big(}
              + 1125\, \* D_{1}^{2}\,
              - 2070\, \* D_{1}^{3}\,
              - 3456\, \* D_{1}^{4}\,
              + 3240\, \* D_{1}^{5}\,
              - 1812\, \* D_{2}^{2}\,
              - 2448\, \* D_{2}^{3}\,
              - 1728\, \* D_{2}^{4}\,
   \nn \\[0mm] & & \mbox{} \vphantom{\Big(}
              + 352\, \* D_{-1}\,
              + 24\, \* (427 + 27\, \* \z3)\, \* D_{1}\,
              - (763 + 648\, \* \z3)\, \* D_{2}\,
              - 12\, \* (802 + 27\, \* \z3)\, \* D_{0}\Big)\,
   \nn \\[0mm] & & \mbox{} \vphantom{\Big(}
          + 4/729\: \* \Big(17370\, \* D_{0}^{4}\,
              - 15012\, \* D_{0}^{5}\,
              - 25992\, \* D_{1}^{4}\,
              + 49464\, \* D_{1}^{5}\,
              - 28512\, \* D_{1}^{6}\,
              - 5280\, \* D_{2}^{3}\,
   \nn \\[0mm] & & \mbox{} \vphantom{\Big(}
              - 3456\, \* D_{2}^{4}\,
              + 13824\, \* D_{2}^{5}\,
              + 128\, \* (31 + 27\, \* \z3)\, \* D_{-1}\,
              - 6\, \* (281 - 9936\, \* \z3)\, \* D_{1}\,
   \nn \\[0mm] & & \mbox{} \vphantom{\Big(}
              + 72\, \* (635 - 18\, \* \z3)\, \* D_{1}^{2}\,
              - 54\, \* (835 + 144\, \* \z3)\, \* D_{0}^{3}\,
              + 24\, \* (959 - 432\, \* \z3)\, \* D_{2}^{2}\,
   \nn \\[0mm] & & \mbox{} \vphantom{\Big(}
              - 6\, \* (1621 - 2592\, \* \z3)\, \* D_{1}^{3}\,
              + 24\, \* (1988 + 459\, \* \z3)\, \* D_{0}^{2}\,
              - 9\, \* (7037 + 3852\, \* \z3)\, \* D_{0}\,
   \nn \\[0mm] & & \mbox{} \vphantom{\Big(}
              + 2\, \* (31649 - 14688\, \* \z3)\, \* D_{2}\Big)\Big\}\,
\nn \\ & & \mbox{} \vphantom{\Big(} \hspace{-5mm}
      + \colourcolour{\ca} \, \*  \Big\{32/27\: \* \Big[
          4\, \* \S(-4)\,
          + \Ssss(1,1,1,1)\,
          - \Sss(1,1,2)\,
          + \Sss(1,2,1)\,
          - \Ss(1,3)\,
          + \Sss(2,1,1)\,
          - \Ss(2,2)\,
          + \Ss(3,1)\,
   \nn \\[0mm] & & \mbox{} \vphantom{\Big(}
          + 3\, \* \S(4)\,\Big]\, \* \npqg 
          - 128/81\: \* \S(-3)\, \* \Big(5\, \* D_{0}\,
              - 7\, \* D_{1}\,
              + 7\, \* D_{2}\Big)\,
          + 64/81\: \* \Big[-\Sss(1,1,1)
                            +\Ss(1,2)
   \nn \\[0mm] & & \mbox{} \vphantom{\Big(}
                            -\Ss(2,1)\Big]\, \* \Big(5\, \* D_{0}\,
              - 10\, \* D_{1}\,
              + 3\, \* D_{1}^{2}\,
              + 10\, \* D_{2}\,
              - 3\, \* D_{2}^{2}\Big)\,
          - 64/81\: \* \S(3)\, \* \Big(5\, \* D_{0}\,
              - 4\, \* D_{1}\,
   \nn \\[0mm] & & \mbox{} \vphantom{\Big(}
              - 3\, \* D_{1}^{2}\,
              + 4\, \* D_{2}\,
              + 3\, \* D_{2}^{2}\Big)\,
          + 16/243\: \* \S(-2)\, \* \Big(38\, \* D_{0}\,
              - 10\, \* D_{1}\,
              + 9\, \* D_{1}^{2}\,
              + 28\, \* D_{2}\Big)\,
   \nn \\[0mm] & & \mbox{} \vphantom{\Big(}
          - 4/243\: \* \Ss(1,1)\, \* \Big(316\, \* D_{0}\,
              - 45\, \* D_{0}^{2}\,
              + 144\, \* D_{0}^{3}\,
              - 641\, \* D_{1}\,
              - 354\, \* D_{1}^{2}\,
              + 349\, \* D_{2}\,
   \nn \\[0mm] & & \mbox{} \vphantom{\Big(}
              + 792\, \* D_{2}^{2}\,
              - 288\, \* D_{2}^{3}\,
              - 104\, \* D_{-1}\Big)\,
          - 4/243\: \* \S(2)\, \* \Big(468\, \* D_{0}\,
              - 45\, \* D_{0}^{2}\,
              + 144\, \* D_{0}^{3}\,
   \nn \\[0mm] & & \mbox{} \vphantom{\Big(}
              - 1659\, \* D_{1}\,
              + 912\, \* D_{1}^{2}\,
              - 576\, \* D_{1}^{3}\,
              + 1277\, \* D_{2}\,
              - 168\, \* D_{2}^{2}\,
              + 288\, \* D_{2}^{3}\,
              - 104\, \* D_{-1}\Big)\,
   \nn \\[0mm] & & \mbox{} \vphantom{\Big(}
          - 2/729\: \* \S(1)\, \* \Big(6354\, \* D_{0}^{2}\,
              - 3258\, \* D_{0}^{3}\,
              + 3456\, \* D_{0}^{4}\,
              + 5298\, \* D_{1}^{2}\,
              + 648\, \* D_{1}^{3}\,
              - 5184\, \* D_{1}^{4}\,
   \nn \\[0mm] & & \mbox{} \vphantom{\Big(}
              + 15408\, \* D_{2}^{2}\,
              + 16992\, \* D_{2}^{3}\,
              - 3456\, \* D_{2}^{4}\,
              - 128\, \* D_{-1}\,
              - 6\, \* (1895 + 864\, \* \zeta_3)\, \* D_{1}\,
   \nn \\[0mm] & & \mbox{} \vphantom{\Big(}
              - 3\, \* (2863 - 864\, \* \zeta_3)\, \* D_{0}\,
              + (17447 + 5184\, \* \zeta_3)\, \* D_{2}\Big)\,
          + 2/243\: \* \Big(554\, \* D_{0}^{3}\,
              + 696\, \* D_{0}^{4}\,
   \nn \\[0mm] & & \mbox{} \vphantom{\Big(}
              + 432\, \* D_{0}^{5}\,
              + 8508\, \* D_{1}^{3}\,
              - 6816\, \* D_{1}^{4}\,
              + 3168\, \* D_{1}^{5}\,
              + 2720\, \* D_{2}^{3}\,
              - 4608\, \* D_{2}^{4}\,
              + 2304\, \* D_{2}^{5}\,
   \nn \\[0mm] & & \mbox{} \vphantom{\Big(}
              - 192\, \* (2 - 3\, \* \zeta_3)\, \* D_{-1}\,
              + 6\, \* (125 + 288\, \* \zeta_3)\, \* D_{1}\,
              - 3\, \* (269 + 912\, \* \zeta_3)\, \* D_{2}\,
   \nn \\[0mm] & & \mbox{} \vphantom{\Big(}
              + 2\, \* (643 - 432\, \* \zeta_3)\, \* D_{0}^{2}\,
              + 8\, \* (653 - 216\, \* \zeta_3)\, \* D_{2}^{2}\,
              - (655 - 432\, \* \zeta_3)\, \* D_{0}\,
   \nn \\[0mm] & & \mbox{} \vphantom{\Big(}
              - 2\, \* (2399 + 864\, \* \zeta_3)\, \* D_{1}^{2}\Big)\Big\}
\eea
and
\bea
\label{ggqnf3}
\lefteqn{\ggq{3}\big|_{\nft}(N) \;=\,
          \colourcolour{\cf} \, \*  \Big\{\,
          - 64/27\: \* \Sss(1,1,1)\, \* \npgq
          + 64/81\: \* \Ss(1,1)\, \* \Big( 8\, \* \npgq
              - 3\, \* D_{1}^{2}\Big)\,
}
   \nn \\[0mm] & & \mbox{} \vphantom{\Big(}
          - 64/81\: \* \S(1)\, \* \Big(
                4\, \* \npgq
              - 8\, \* D_{1}^{2}\,
              + 3\, \* D_{1}^{3}\Big)\,
          - 64/81\: \* \Big(6\, \* \npgq\, \* \z3
              + 4\, \* D_{1}^{2}\,
              - 8\, \* D_{1}^{3}\,
              + 3\, \* D_{1}^{4}\Big)
          \Big\}\,
\:\: . \quad
\eea
Finally the corresponding contribution to the gluon-gluon anomalous 
dimension reads
\bea
\label{gggnf3}
\lefteqn{\ggg{3}\big|_{\nft}(N) \;=\,
          \colourcolour{\cf} \, \*  \Big\{\,
          64/27\: \* \Big(3\, \* D_{0}\,
              - 6\, \* D_{0}^{2}\,
              - 3\, \* D_{1}\,
              - 6\, \* D_{1}^{2}\,
              - 4\, \* D_{2}\,
              + 4\, \* D_{-1}\Big)\, \* \Big[\Sss(1,1,1)\,
              - \Ss(1,2)\,
}
   \nn \\[0mm] & & \mbox{} \vphantom{\Big(}
              - \Ss(2,1)\,
              + \S(3)/2\Big]\,
           + 64/81\: \* \Ss(1,1)\, \* \Big(57\, \* D_{0}\,
              + 21\, \* D_{0}^{2}\,
              + 18\, \* D_{0}^{3}\,
              - 39\, \* D_{1}\,
              + 12\, \* D_{1}^{2}\,
              + 20\, \* D_{2}\,
   \nn \\[0mm] & & \mbox{} \vphantom{\Big(}
              - 38\, \* D_{-1}\Big)\,
          - 32/81\: \* \S(2)\, \* \Big(42\, \* D_{0}\,
              + 69\, \* D_{0}^{2}\,
              + 18\, \* D_{0}^{3}\,
              - 42\, \* D_{1}\,
              + 69\, \* D_{1}^{2}\,
              - 18\, \* D_{1}^{3}\,
   \nn \\[0mm] & & \mbox{} \vphantom{\Big(}
              + 70\, \* D_{2}\,
              - 70\, \* D_{-1}\Big)\,
          - 32/243\: \* \S(1)\, \* \Big(429\, \* D_{0}\,
              + 276\, \* D_{0}^{2}\,
              + 207\, \* D_{0}^{3}\,
              + 54\, \* D_{0}^{4}\,
              - 33\, \* D_{1}\,
   \nn \\[0mm] & & \mbox{} \vphantom{\Big(}
              - 30\, \* D_{1}^{2}\,
              + 135\, \* D_{1}^{3}\,
              - 54\, \* D_{1}^{4}\,
              - 26\, \* D_{2}\,
              - 370\, \* D_{-1}\Big)\,
          - 2/243\: \* \Big(77\,
              - 3360\, \* D_{0}^{3}\,
   \nn \\[0mm] & & \mbox{} \vphantom{\Big(}
              - 1656\, \* D_{0}^{4}\,
              - 432\, \* D_{0}^{5}\,
              - 3840\, \* D_{1}^{3}\,
              + 3816\, \* D_{1}^{4}\,
              - 1296\, \* D_{1}^{5}\,
              - 1296\, \* (3 + \z3)\, \* D_{1}\,
   \nn \\[0mm] & & \mbox{} \vphantom{\Big(}
              - 432\, \* (11 - 3\, \* \z3)\, \* D_{0}\,
              + 96\, \* (43 - 18\, \* \z3)\, \* D_{2}\,
              + 96\, \* (47 + 18\, \* \z3)\, \* D_{-1}\,
   \nn \\[0mm] & & \mbox{} \vphantom{\Big(}
              - 24\, \* (179 + 108\, \* \z3)\, \* D_{0}^{2}\,
              + 24\, \* (193 - 108\, \* \z3)\, \* D_{1}^{2}\Big)\,
          \Big\}\,
\nn \\ & & \mbox{} \vphantom{\Big(} \hspace{-5mm}         
            + \colourcolour{\ca} \, \*  \Big\{\,
            4/81\: \* \Big[-2\, \* \Ss(1,1) + \S(2)\Big]\, \* \Big(33\, \* D_{0}\,
              + 48\, \* D_{0}^{2}\,
              - 33\, \* D_{1}\,
              + 48\, \* D_{1}^{2}\,
              + 52\, \* D_{2}\,
              - 52\, \* D_{-1}\Big)\,
   \nn \\[0mm] & & \mbox{} \vphantom{\Big(}
          + 4/243\: \* \S(1)\, \* \Big(480\, \* D_{0}\,
              + 456\, \* D_{0}^{2}\,
              + 144\, \* D_{0}^{3}\,
              - 480\, \* D_{1}\,
              + 456\, \* D_{1}^{2}\,
              - 144\, \* D_{1}^{3}\,
              + 527\, \* D_{2}\,
   \nn \\[0mm] & & \mbox{} \vphantom{\Big(}
              - 527\, \* D_{-1}\,
              - 24\, \* (1 - 6\, \* \z3)\Big)\,
          - 1/243\: \* \Big(5\,
              + 1380\, \* D_{0}^{2}\,
              + 912\, \* D_{0}^{3}\,
              + 288\, \* D_{0}^{4}\,
   \nn \\[0mm] & & \mbox{} \vphantom{\Big(}
              + 1380\, \* D_{1}^{2}\,
              - 912\, \* D_{1}^{3}\,
              + 288\, \* D_{1}^{4}\,
              + 6\, \* (229 - 96\, \* \z3)\, \* D_{0}\,
              - 6\, \* (229 - 96\, \* \z3)\, \* D_{1}\,
   \nn \\[0mm] & & \mbox{} \vphantom{\Big(}
              + 4\, \* (331 - 144\, \* \z3)\, \* D_{2}\,
              - 4\, \* (331 - 144\, \* \z3)\, \* D_{-1}\Big)
          \Big\}\,
\:\: .
\eea
The $C_A$ part of Eq.~(\ref{gggnf3}), which is a non-singlet$\,$-type
quantity and hence could be written in a more compact manner in terms
of the quantities in Eq.~(\ref{EtaNu}), has been obtained already in
Ref.~\cite{LargeNf3}. Its~leading large-$N$ coefficient is related to
that in Eq.~(\ref{A4nf23}) by the `Casimir scaling' $C_A/C_F$.
Moreover two linear combinations of Eq.~(\ref{ggqnf3}) with 
Eq.~(\ref{gpsnf3}) and the $C_F$ part of Eq.~(\ref{gggnf3}) were 
derived in Ref.~\cite{LargeNf2,LargeNf3}; our results agree also with 
those findings. Eq.~(\ref{gqgnf3}) is entirely new.

\begin{figure}[p]
\vspace{-4mm}
\centerline{\epsfig{file=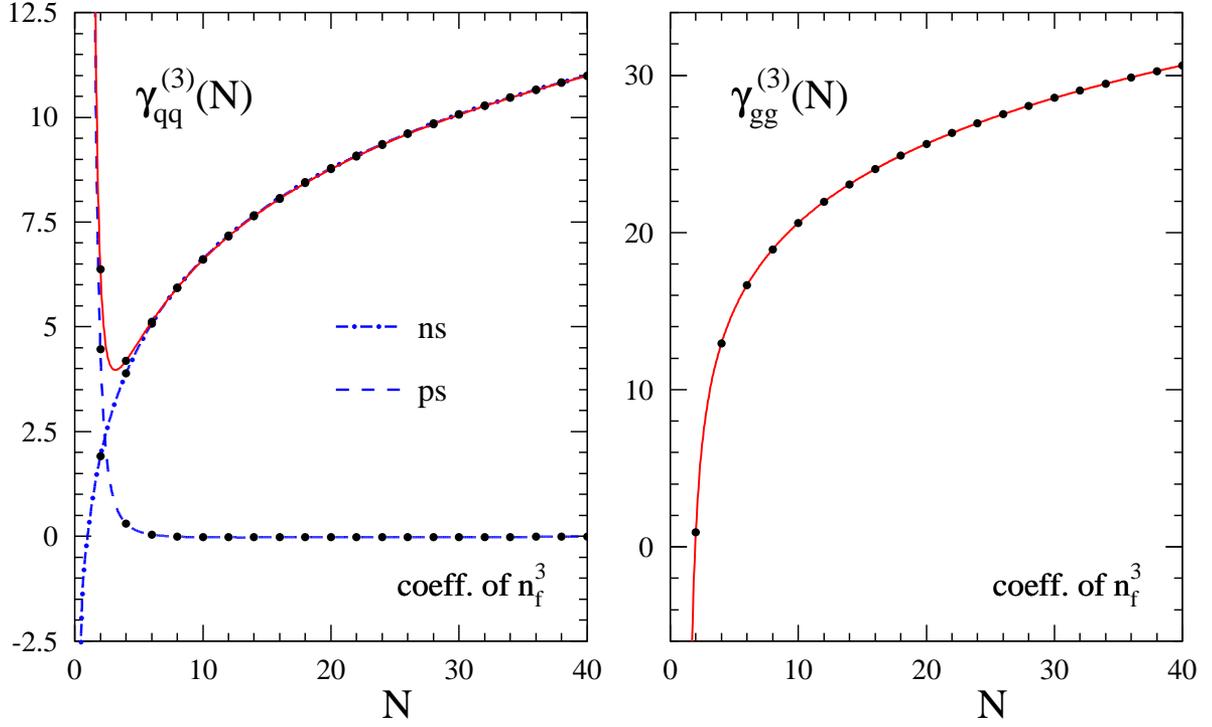,width=16.0cm,angle=0}}
\vspace{-2mm}
\caption{ \label{fig:gdiagnf3} \small
 The $\nft$ parts of the `diagonal' quark-quark and gluon-gluon four-loop 
 anomalous dimensions. The analytically calculated even-$N$ moments are
 shown together with their continuation calculated via a numerical Mellin 
 transformation of the corresponding $x$-space expressions using the
 program of Ref.~\cite{HPLnum}. 
 For the quark-quark case the non-singlet and pure-singlet contributions 
 are displayed separately.}
\vspace{-1mm}
\end{figure}

\begin{figure}[p]
\vspace{-2mm}
\centerline{\epsfig{file=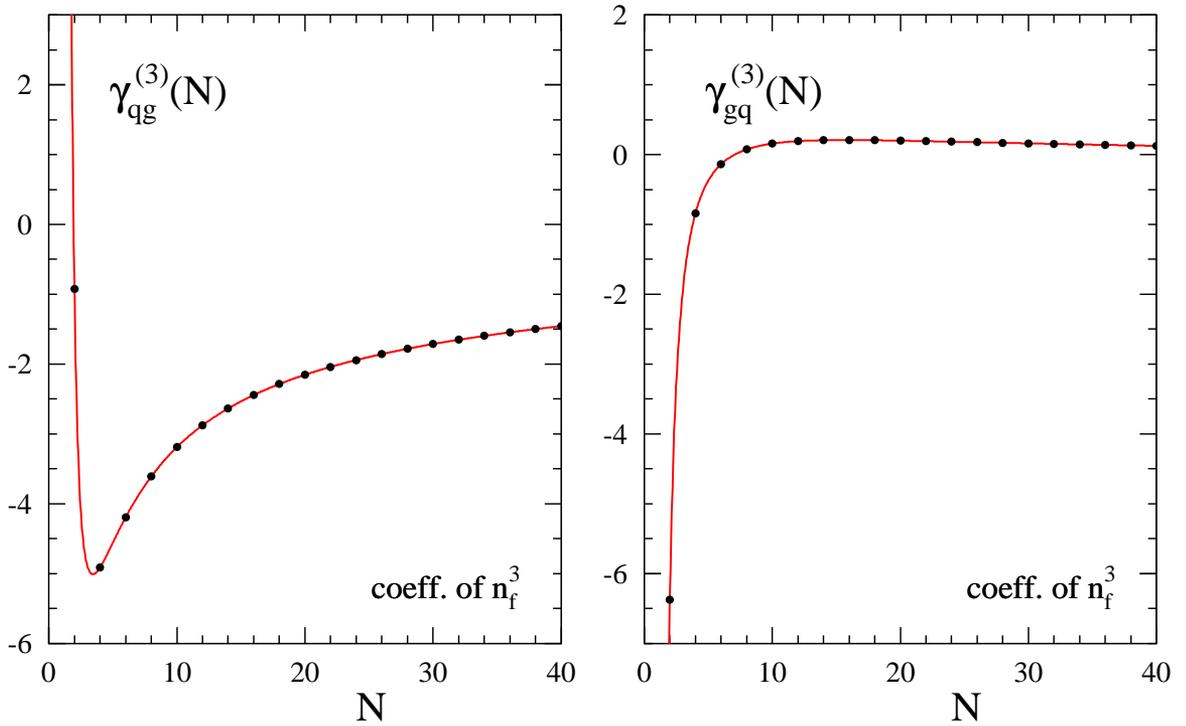,width=16.0cm,angle=0}}
\vspace{-2mm}
\caption{ \label{fig:goffdnf3} \small
 As Figure~\ref{fig:gdiagnf3}, but for the `off-diagonal' gluon-quark 
 and quark-gluon anomalous dimensions.}
\vspace{-1mm}
\end{figure}

The results (\ref{gns3nf3}), (\ref{gpsnf3}) and (\ref{gqgnf3}) -- 
(\ref{gggnf3}) and their  continuations to non-integer $N$ are illustrated 
in Figs.~\ref{fig:gdiagnf3} and \ref{fig:goffdnf3} for the normalization
specified for their $x$-space counterparts in Eq.~(\ref{Pexp}).


\newpage

%
\setcounter{equation}{0}
\section{ Results in $x\:\!$-$\:\!$space}
\label{sec:xres}

The four-loop splitting functions $P_{\rm ik}^{\,(3)}(x)$ are obtained 
from the above $N$-space results by an inverse Mellin transformation which 
expresses these functions in terms of harmonic polylogarithms.
This transformation can be performed by a completely algebraic procedure 
\cite{Hpols,MV99} based on the fact that harmonic sums 
occur as coefficients of the Taylor expansion of harmonic polylogarithms.

Before we present our results, we recall the basic definitions \cite{Hpols}:
The lowest-weight ($w = 1$) functions $H_m(x)$ are given by
\beq
\label{HPL1}
  H_0(x)       \; = \; \ln x \:\: , \quad\quad
  H_{\pm 1}(x) \; = \; \mp \, \ln (1 \mp x) \:\: .
\eeq
The higher-weight ($w \geq 2$) functions are recursively defined as
\beq
\label{HPL2}
  H_{m_1,...,m_w}(x) \; = \;
    \left\{ \begin{array}{cl}
    \displaystyle{ \frac{1}{w!}\,\ln^{\,w\!} x \:\: ,}
       & \quad {\rm if} \:\:\: m^{}_1,...,m^{}_w \,=\, 0,\ldots ,0 
\\[2ex]
    \displaystyle{ \int_0^x \! dz\: f_{m_1}(z) \, H_{m_2,...,m_w}(z)
       \:\: , } & \quad {\rm otherwise}
    \end{array} \right.
\eeq
with
\beq
\label{HPLf}
  f_0(x)       \; = \; \frac{1}{x} \:\: , \quad\quad
  f_{\pm 1}(x) \; = \; \frac{1}{1 \mp x} \;\; .
\eeq
For chains of indices `zero' we employ the abbreviated notation
\beq
\label{HPP-abbr}
  H_{{\footnotesize \underbrace{0,\ldots ,0}_{\scriptstyle m} },\,
  \pm 1,\, {\footnotesize \underbrace{0,\ldots ,0}_{\scriptstyle n} },
  \, \pm 1,\, \ldots}(x) \; = \; H_{\pm (m+1),\,\pm (n+1),\, \ldots}(x)
\eeq
and suppress the argument $x$ in all results below. 

The splitting functions for the quark$\,\pm\,$antiquark flavour differences
in Eq.~(\ref{qns}) are expressed in a decomposition analogous to the first 
line of Eq.~(\ref{gns3Nf2}),
\beq
\label{Pns3dec}
 P_{\rm ns}^{\,(3)\pm}(x) \big|_\nfs \:\: = \:\: \cf \nfs \left\{ 
  2\:\! \cf\, \widetilde{A}^{\,(3)} 
  \:+\: ( \ca - 2\:\!\cf ) \widetilde{B}_\pm^{\,(3)} \right\}
\eeq
with
\bea
\label{pns3Anf2}
\lefteqn{\widetilde{A}^{\,(3)}(x) \;=\;
 - \: \frct{16}{9}\, \* \Big\{
       \pqq{x}\, \* \Big(6\, \* \Hhhh(0,0,0,0)\,
              - \Hhhh(1,0,0,0)\,
              - 2\, \* \Hh(1,3)\,
              + 2\, \* \Hhh(2,0,0)\,
              + 4\, \* \Hh(3,0)\,
              + 5\, \* \H(4)\,
\qquad \qquad \qquad}
   \nn \\[0mm] & & \mbox{} \vphantom{\Big(}
              + \frct{53}{4}\: \* \Hhh(0,0,0)\,
              + \frct{10}{3}\: \* \Hhh(1,0,0)\,
              + \frct{20}{3}\: \* \Hh(2,0)\,
              + 10\, \* \H(3)\,
              + \frct{287}{18}\: \* \Hh(0,0)\,
              - 5\, \* \Hh(0,0)\, \* \zeta_2\,
              + \frct{19}{9}\: \* \Hh(1,0)\,
   \nn \\[0mm] & & \mbox{} \vphantom{\Big(}
              + 2\, \* \Hh(1,0)\, \* \zeta_2\,
              + \frct{19}{9}\: \* \H(2)\,
              + \frct{1259}{72}\: \* \H(0)\,
              + 6\, \* \H(0)\, \* \zeta_3\,
              - 10\, \* \H(0)\, \* \zeta_2\,
              + 2\, \* \H(1)\, \* \zeta_3\,
              - \frct{19}{9}\: \* \zeta_2\,
              + \frct{40}{3}\: \* \zeta_3\,
   \nn \\[0mm] & & \mbox{} \vphantom{\Big(}
              - \frct{7}{2}\: \* \zeta_4\,
              + \frct{2119}{288}\Big)\,
       + (1-x)\, \* \Big(2\, \* \Hhh(1,0,0)\,
              + \frct{29}{3}\: \* \Hh(1,0)\,
              + \frct{23}{3}\: \* \H(1)\Big)\,
       + x\, \* \Big( - 3\, \* \Hhhh(0,0,0,0)\,
   \nn \\[0mm] & & \mbox{} \vphantom{\Big(}
              - \frct{37}{4}\: \* \Hhh(0,0,0)\,
              - 3\, \* \Hh(2,0)\,
              - 3\, \* \H(3)\,
              - \frct{188}{9}\: \* \Hh(0,0)\,
              - \frct{35}{6}\: \* \H(2)\,
              - \frct{2539}{48}\: \* \H(0)\,
              + 3\, \* \H(0)\, \* \zeta_2\,
              + \frct{35}{6}\: \* \zeta_2\,
   \nn \\[0mm] & & \mbox{} \vphantom{\Big(}
              - 9\, \* \zeta_3\,
              - \frct{5729}{72}\Big)\,
              - 3\, \* \Hhhh(0,0,0,0)\,
              + \frct{7}{4}\: \* \Hhh(0,0,0)\,
              + 5\, \* \Hh(2,0)\,
              + 7\, \* \H(3)\,
              + \frct{232}{9}\: \* \Hh(0,0)\,
              + \frct{27}{2}\: \* \H(2)\,
   \nn \\[0mm] & & \mbox{} \vphantom{\Big(}
              + \frct{8911}{144}\: \* \H(0)\,
              - 7\, \* \H(0)\, \* \zeta_2\,
              - \frct{27}{2}\: \* \zeta_2\,
              + 9\, \* \zeta_3\,
              + \frct{5729}{72}\,
       + \delta(1-x)\, \* \Big(
              - \frct{127}{32}
              + \frct{1259}{36}\: \* \zeta_2\,
   \nn \\[0mm] & & \mbox{} \vphantom{\Big(}
              - \frct{233}{12}\: \* \zeta_3\,
              - \frct{323}{12}\: \* \zeta_4\,
              + 10\, \* \zeta_3\, \*  \zeta_2\,
              + 6\, \* \zeta_5\,
              \Big) \Big\}
\eea
and
\bea
\label{pns3Bnf2}
\lefteqn{\widetilde{B}_+^{\:(3)}(x) \;=\;
 -\: \frct{32}{9}\: \* \Big\{
       \pqq{x}\, \* \Big( - \Hh(-3,0)\,
              + \frct{3}{2}\: \* \Hhhh(0,0,0,0)\,
              - 2\, \* \Hhh(1,-2,0)\,
              - 4\, \* \Hhhh(1,0,0,0)\,
              - \Hh(1,3)\,
}
   \nn \\[0mm] & & \mbox{} \vphantom{\Big(}
              - \frct{3}{2}\: \* \Hhh(2,0,0)\,
              + \frct{1}{2}\: \* \Hh(3,0)\,
              + \H(4)\,
              + \frct{73}{12}\: \* \Hhh(0,0,0)\,
              - 5\, \* \Hhh(1,0,0)\,
              + \frct{5}{3}\: \* \H(3)\,
              + \frct{619}{72}\: \* \Hh(0,0)\,
              - 2\, \* \Hh(0,0)\, \* \zeta_2\,
   \nn \\[0.5mm] & & \mbox{} \vphantom{\Big(}
              + \frct{1585}{288}\: \* \H(0)\,
              + \frct{7}{2}\: \* \H(0)\, \* \zeta_3\,
              - \frct{10}{3}\: \* \H(0)\, \* \zeta_2\,
              - 2\, \* \H(1)\, \* \zeta_3\,
              - \frct{19}{18}\: \* \zeta_2\,
              + 15\, \* \zeta_3\,
              - 6\, \* \zeta_4\,
              + \frct{923}{576}\Big)\,
   \nn \\[0mm] & & \mbox{} \vphantom{\Big(}
       + \frct{1}{2}\, \* \pqq{-x}\, \* \Big(
                2\, \* \Hh(-3,0)\,
              + 2\, \* \Hhh(-2,0,0)\,
              - 4\, \* \Hh(-2,2)\,
              + 2\, \* \Hhhh(-1,0,0,0)\,
              + 8\, \* \Hhh(-1,2,1)\,
              - 4\, \* \Hh(-1,3)\,
   \nn \\[0mm] & & \mbox{} \vphantom{\Big(}
              - 3\, \* \Hhhh(0,0,0,0)\,
              - 4\, \* \Hh(3,1)\,
              + 4\, \* \H(4)\,
              + \frct{20}{3}\: \* \Hh(-2,0)\,
              + \frct{20}{3}\: \* \Hhh(-1,0,0)\,
              - \frct{40}{3}\: \* \Hh(-1,2)\,
              - \frct{20}{3}\: \* \Hhh(0,0,0)\,
   \nn \\[0mm] & & \mbox{} \vphantom{\Big(}
              + \frct{20}{3}\: \* \H(3)\,
              + 4\, \* \H(-2)\, \* \zeta_2\,
              + \frct{38}{9}\: \* \Hh(-1,0)\,
              + 2\, \* \Hh(-1,0)\, \* \zeta_2\,
              - \frct{19}{9}\: \* \Hh(0,0)\,
              - 2\, \* \Hh(0,0)\, \* \zeta_2\,
   \nn \\[0.5mm] & & \mbox{} \vphantom{\Big(}
              - 4\, \* \H(-1)\, \* \zeta_3\,
              + \frct{40}{3}\: \* \H(-1)\, \* \zeta_2\,
              - \H(0)\, \* \zeta_3\,
              - \frct{10}{3}\: \* \H(0)\, \* \zeta_2\,
              + \frct{19}{9}\: \* \zeta_2\,
              - 10\, \* \zeta_3\,
              + \frct{1}{2}\: \* \zeta_4\Big)\,
   \nn \\[0mm] & & \mbox{} \vphantom{\Big(}
       + (1-x)\, \* \Big( - 3\, \* \Hhh(1,0,0)\,
              + \Hh(1,0)\,
              - 4\, \* \Hh(1,1)\,
              + 14\, \* \H(1)\Big)\,
       + (1+x)\, \* \Big(2\, \* \Hhh(-1,0,0)\,
              - 4\, \* \Hh(-1,2)\,
   \nn \\[0mm] & & \mbox{} \vphantom{\Big(}
              + \frct{1}{2}\: \* \Hh(2,0)\,
              - 2\, \* \Hh(2,1)\,
              + \frct{17}{3}\: \* \Hh(-1,0)\,
              + 4\, \* \H(-1)\, \* \zeta_2\Big)\,
       + x\, \* \Big(3\, \* \Hh(-2,0)\,
              + \frct{7}{2}\: \* \H(3)\,
              - \frct{211}{24}\: \* \Hh(0,0)\,
   \nn \\[0.5mm] & & \mbox{} \vphantom{\Big(}
              + \frct{31}{4}\: \* \H(2)\,
              - \frct{139}{4}\: \* \H(0)\,
              - \frct{1}{2}\: \* \H(0)\, \* \zeta_2\,
              - \frct{25}{12}\: \* \zeta_2\,
              - 10\, \* \zeta_3\,
              - \frct{1187}{48}\Big)\,
              + \Hh(-2,0)\,
              + \frct{9}{2}\: \* \H(3)\,
   \nn \\[0.5mm] & & \mbox{} \vphantom{\Big(}
              + \frct{47}{8}\: \* \Hh(0,0)\,
              + \frct{39}{4}\: \* \H(2)\,
              + \frct{83}{4}\: \* \H(0)\,
              - \frct{9}{2}\: \* \H(0)\, \* \zeta_2\,
              - \frct{39}{4}\: \* \zeta_2\,
              + 6\, \* \zeta_3\,
              + \frct{1187}{48}\,
   \nn \\[0.5mm] & & \mbox{} \vphantom{\Big(}
       + \delta(1-x)\, \* \Big(
              - \frct{193}{192}
              + \frct{1585}{144}\: \* \zeta_2\,
              - 10\, \* \zeta_3\,
              - \frct{5}{2}\: \* \zeta_4\,
              + \frct{15}{2}\: \* \zeta_3\, \*  \zeta_2\,
              - \frct{11}{4}\: \* \zeta_5\,
              \Big) \Big\}
\:\: ,
\eea
where we have used the abbreviation 
\beq
\label{pqq}
   \pqq{x} \;=\; 2\, (1-x)^{-1} - 1 - x
\:\: .
\eeq
All divergences for $x \to 1$ are to be read as plus-distributions%
. 
The second contribution to 
$P_{\rm ns}^{\,(3)-}(x)$ in Eq.~(\ref{Pns3dec}) can be expressed via
\beq
  \widetilde{B}_-^{\:(3)}(x) \:\:=\:\: \widetilde{B}_+^{\:(3)}(x) 
  \:+\: \delta \widetilde{B}^{\:(3)}(x)
\eeq
and
\bea
\label{pns3dBnf2}
\lefteqn{\delta \widetilde{B}^{\:(3)}(x) \;=\;
        - \, \frct{32}{9}\: \* \Big\{
        \pqq{-x}\, \* \Big(
        2\, \* \H(-3,0) 
        + 2\, \* \H(-2,0,0)
        - 4\, \* \H(-2,2)
        + 2\, \* \H(-1,0,0,0)
}
   \nn \\[0mm] & & \mbox{} \vphantom{\Big(}
        + 8\, \* \H(-1,2,1)
        - 4\, \* \H(-1,3)
        - 3\, \* \H(0,0,0,0)
        - 4\, \* \H(3,1)
        + 4\, \* \H(4)
        + \frct{20}{3}\: \* \H(-2,0)
        + \frct{20}{3}\: \* \H(-1,0,0)
   \nn \\[0mm] & & \mbox{} \vphantom{\Big(}
        - \frct{40}{3}\: \* \H(-1,2)
        - \frct{20}{3}\: \* \H(0,0,0)
        + \frct{20}{3}\: \* \H(3)
        + 4\, \* \H(-2)\, \* \z2
        + \frct{38}{9}\: \* \H(-1,0)
        + 2\, \* \H(-1,0)\, \* \z2
   \nn \\[0mm] & & \mbox{} \vphantom{\Big(}
        - \frct{19}{9}\: \* \H(0,0)
        - 2\, \* \H(0,0)\, \* z2
        + \frct{40}{3}\: \* \H(-1)\, \* \z2
        - 4\, \* \H(-1)\, \* \z3
        - \frct{10}{3}\: \* \H(0)\, \* \z2
        - H(0)\, \* \z3
   \nn \\[0mm] & & \mbox{} \vphantom{\Big(}
        + \frct{19}{9}\: \* \z2
        - 10\, \* \z3
        + \frct{1}{2}\: \* \z4
        \Big)
        - (1-x)\, \* \Big(
        + 8\, \* \H(1,1)
        - \frct{61}{3}\: \* \H(1)
        + \frct{277}{18}
        \Big)
   \nn \\[0mm] & & \mbox{} \vphantom{\Big(}
        + (1+x)\, \* \Big(
        4\, \* \H(-2,0)
        + 4\, \* \H(-1,0,0)
        - 8\, \* \H(-1,2)
        - 5\, \* \H(0,0,0)
        - 4\, \* \H(2,1)
        + 6\, \* \H(3)
        - \frct{29}{2}\: \* \H(0,0)
   \nn \\[0mm] & & \mbox{} \vphantom{\Big(}
        + \frct{46}{3}\: \* \H(-1,0)
        + \frct{41}{3}\: \* \H(2)
        + 8\, \* \H(-1)\, \* \z2
        - \frct{151}{9}\: \* \H(0)
        - 3\, \* \H(0)\, \* \z2
        - 4\, \* \z3
        \Big)
        - (4+8\, \* x)\, \* \z2 
        \Big\}
\:\: .
\eea
  
The inverse Mellin transform of Eq.~(\ref{eqn:S}), up to the conventional
minus sign between the anomalous dimensions and splitting functions, is
given by the rather lengthy expression
\bea
\label{pns3Snf2}
\lefteqn{ P_{\,\rm ns}^{\,(3)\rm s}\big|_{\nfs \, \dabctnc} \;=\;
\frct{128}{3}\, \* \Big\{
        (\frct{1}{x}-x^{2})\, \* \Big( - \frct{16}{3}\: \* \Hhh(1,-2,0)\,
              + \frct{16}{3}\: \* \Hhhh(1,0,0,0)\,
              + \frct{8}{3}\: \* \Hhhh(1,1,0,0)\,
              + \frct{8}{3}\: \* \Hh(1,3)\,
}
   \nn \\[1mm] & & \mbox{} \vphantom{\Big(}
              - \frct{20}{3}\: \* \Hh(1,0)\, \* \zeta_2\,
              - \frct{16}{3}\: \* \Hh(1,1)\, \* \zeta_2\,
              - \frct{44}{3}\: \* \H(1)\, \* \zeta_3\,
              + \frct{40}{9}\: \* \H(1)\, \* \zeta_2\Big)\,
       + (\frct{1}{x}+x^{2})\, \* \Big( - \frct{8}{3}\: \* \Hhh(-1,-2,0)\,
   \nn \\[0mm] & & \mbox{} \vphantom{\Big(}
              + \frct{32}{3}\: \* \Hhhh(-1,-1,-1,0)\,
              - \frct{16}{3}\: \* \Hhhh(-1,-1,0,0)\,
              - \frct{32}{3}\: \* \Hhh(-1,-1,2)\,
              - \frct{8}{3}\: \* \Hhhh(-1,0,0,0)\,
              + \frct{8}{3}\: \* \Hhh(-1,2,0)\,
   \nn \\[0mm] & & \mbox{} \vphantom{\Big(}
              + \frct{32}{3}\: \* \Hhh(-1,2,1)\,
              + \frct{16}{3}\: \* \Hh(-1,3)\,
              + \frct{80}{9}\: \* \Hhh(-1,-1,0)\,
              - \frct{80}{9}\: \* \Hhh(-1,0,0)\,
              - \frct{80}{9}\: \* \Hh(-1,2)\,
              + 16\, \* \Hh(-1,-1)\, \* \zeta_2\,
   \nn \\[0mm] & & \mbox{} \vphantom{\Big(}
              - \frct{28}{3}\: \* \Hh(-1,0)\, \* \zeta_2\,
              - \frct{56}{3}\: \* \H(-1)\, \* \zeta_3\,
              + \frct{40}{3}\: \* \H(-1)\, \* \zeta_2\Big)\,
       + (1-x)\, \* \Big(2\, \* \Hhh(-3,0,0)\,
              - 4\, \* \Hh(-3,2)\,
   \nn \\[0mm] & & \mbox{} \vphantom{\Big(}
              + 4\, \* \Hhh(-2,-2,0)\,
              - 16\, \* \Hhhh(-2,-1,-1,0)\,
              + 4\, \* \Hhhh(-2,-1,0,0)\,
              + 8\, \* \Hhh(-2,-1,2)\,
              + 2\, \* \Hhhh(-2,0,0,0)\,
              - 4\, \* \Hh(-2,3)\,
   \nn \\[0mm] & & \mbox{} \vphantom{\Big(}
              + 13\, \* \Hhh(-2,0,0)\,
              + 8\, \* \Hhh(1,-2,0)\,
              + \Hhhh(1,0,0,0)\,
              - \Hhhh(1,1,0,0)\,
              + 2\, \* \Hh(1,3)\,
              + 4\, \* \H(-3)\, \* \zeta_2\,
              - 16\, \* \Hh(-2,-1)\, \* \zeta_2\,
   \nn \\[0mm] & & \mbox{} \vphantom{\Big(}
              + 6\, \* \Hh(-2,0)\, \* \zeta_2\,
              + \frct{77}{6}\: \* \Hhh(1,0,0)\,
              + 14\, \* \H(-2)\, \* \zeta_3\,
              + \frct{91}{6}\: \* \Hh(1,0)\,
              + \Hh(1,0)\, \* \zeta_2\,
              + \frct{182}{3}\: \* \Hh(1,1)\,
              - 4\, \* \Hh(1,1)\, \* \zeta_2\,
   \nn \\[0mm] & & \mbox{} \vphantom{\Big(}
              - \frct{131}{36}\: \* \H(1)\,
              + 10\, \* \H(1)\, \* \zeta_3\,
              + \frct{16}{3}\: \* \H(1)\, \* \zeta_2\Big)\,
       + (1+x)\, \* \Big(6\, \* \Hhhh(2,0,0,0)\,
              + 2\, \* \Hhhh(2,1,0,0)\,
              + 4\, \* \Hh(2,3)\,
   \nn \\[0mm] & & \mbox{} \vphantom{\Big(}
              + 3\, \* \Hhh(3,0,0)\,
              + 3\, \* \Hh(4,0)\,
              + 12\, \* \Hh(4,1)\,
              + 2\, \* \Hhh(-1,-2,0)\,
              - 8\, \* \Hhhh(-1,-1,-1,0)\,
              - 2\, \* \Hhhh(-1,-1,0,0)\,
   \nn \\[0mm] & & \mbox{} \vphantom{\Big(}
              - 4\, \* \Hhh(-1,-1,2)\,
              - \Hhhh(-1,0,0,0)\,
              + 4\, \* \Hhh(-1,2,0)\,
              + 16\, \* \Hhh(-1,2,1)\,
              + 2\, \* \Hh(-1,3)\,
              - \frct{32}{3}\: \* \Hhh(-1,-1,0)\,
   \nn \\[0mm] & & \mbox{} \vphantom{\Big(}
              + \frct{7}{3}\: \* \Hhh(-1,0,0)\,
              - \frct{70}{3}\: \* \Hh(-1,2)\,
              + \frct{41}{6}\: \* \Hh(2,0)\,
              - 6\, \* \Hh(2,0)\, \* \zeta_2\,
              + \frct{82}{3}\: \* \Hh(2,1)\,
              - 8\, \* \Hh(2,1)\, \* \zeta_2\,
              + \frct{155}{18}\: \* \Hh(-1,0)\,
   \nn \\[0mm] & & \mbox{} \vphantom{\Big(}
              - 5\, \* \Hh(-1,0)\, \* \zeta_2\,
              - 8\, \* \H(2)\, \* \zeta_3\,
              - 7\, \* \H(-1)\, \* \zeta_3\,
              + 18\, \* \H(-1)\, \* \zeta_2\Big)\,
       + x\, \* \Big( - 6\, \* \H(5)\,
              - 16\, \* \Hh(-3,0)\,
   \nn \\[0mm] & & \mbox{} \vphantom{\Big(}
              + 32\, \* \Hhh(-2,-1,0)\,
              - 6\, \* \Hh(-2,2)\,
              + \frct{7}{2}\: \* \Hhh(2,0,0)\,
              - \frct{9}{2}\: \* \Hh(3,0)\,
              - 18\, \* \Hh(3,1)\,
              - 15\, \* \H(4)\,
              + \frct{64}{3}\: \* \Hh(-2,0)\,
   \nn \\[0mm] & & \mbox{} \vphantom{\Big(}
              - 12\, \* \Hhh(0,0,0)\,
              + 6\, \* \Hhh(0,0,0)\, \* \zeta_2\,
              + \frct{85}{3}\: \* \H(3)\,
              + 22\, \* \H(-2)\, \* \zeta_2\,
              - \frct{250}{9}\: \* \Hh(0,0)\,
              + 15\, \* \Hh(0,0)\, \* \zeta_2\,
              - \frct{382}{9}\: \* \H(2)\,
   \nn \\[0mm] & & \mbox{} \vphantom{\Big(}
              + 16\, \* \H(2)\, \* \zeta_2\,
              - \frct{41}{4}\: \* \H(0)\,
              + \frct{25}{2}\: \* \H(0)\, \* \zeta_4\,
              - \frct{85}{3}\: \* \H(0)\, \* \zeta_2\,
              + 14\, \* \zeta_3\, \* \zeta_2\,
              + \frct{382}{9}\: \* \zeta_2\,
              - 10\, \* \zeta_3\,
              - \frct{275}{4}\: \* \zeta_4\,
   \nn \\[0mm] & & \mbox{} \vphantom{\Big(}
              + \frct{850}{9}\Big)\,
       + x^{2}\, \* \Big(\frct{8}{3}\: \* \Hh(-3,0)\,
              - \frct{16}{3}\: \* \Hhh(-2,-1,0)\,
              + \frct{16}{3}\: \* \Hh(-2,2)\,
              + \frct{8}{3}\: \* \Hhhh(0,0,0,0)\,
              - \frct{8}{3}\: \* \Hhh(2,0,0)\,
   \nn \\[0mm] & & \mbox{} \vphantom{\Big(}
              - \frct{8}{3}\: \* \Hh(3,0)\,
              - \frct{32}{3}\: \* \Hh(3,1)\,
              - \frct{8}{3}\: \* \H(4)\,
              - \frct{80}{9}\: \* \Hh(-2,0)\,
              + \frct{80}{9}\: \* \Hhh(0,0,0)\,
              + \frct{80}{9}\: \* \H(3)\,
              - 8\, \* \H(-2)\, \* \zeta_2\,
   \nn \\[0mm] & & \mbox{} \vphantom{\Big(}
              + \frct{16}{3}\: \* \Hh(0,0)\, \* \zeta_2\,
              + \frct{8}{3}\: \* \H(2)\, \* \zeta_2\,
              + 20\, \* \H(0)\, \* \zeta_3\,
              - \frct{160}{9}\: \* \H(0)\, \* \zeta_2\,
              - \frct{200}{9}\: \* \zeta_3\,
              + \frct{19}{3}\: \* \zeta_4\Big)\,
              - 4\, \* \Hhhhh(0,0,0,0,0)\,
   \nn \\[0mm] & & \mbox{} \vphantom{\Big(}
              - 2\, \* \H(5)\,
              + 14\, \* \Hh(-3,0)\,
              - 28\, \* \Hhh(-2,-1,0)\,
              + 2\, \* \Hh(-2,2)\,
              - 12\, \* \Hhhh(0,0,0,0)\,
              + \frct{11}{2}\: \* \Hhh(2,0,0)\,
              + \frct{1}{2}\: \* \Hh(3,0)\,
   \nn \\[0mm] & & \mbox{} \vphantom{\Big(}
              + 2\, \* \Hh(3,1)\,
              - 22\, \* \H(4)\,
              - 2\, \* \Hh(-2,0)\,
              - \frct{26}{3}\: \* \Hhh(0,0,0)\,
              + 2\, \* \Hhh(0,0,0)\, \* \zeta_2\,
              - 5\, \* \H(3)\,
              - 16\, \* \H(-2)\, \* \zeta_2\,
   \nn \\[0mm] & & \mbox{} \vphantom{\Big(}
              - \frct{400}{9}\: \* \Hh(0,0)\,
              - 12\, \* \Hh(0,0)\, \* \zeta_3\,
              + 36\, \* \Hh(0,0)\, \* \zeta_2\,
              - \frct{109}{9}\: \* \H(2)\,
              + 14\, \* \H(2)\, \* \zeta_2\,
              - \frct{725}{9}\: \* \H(0)\,
              - \frct{5}{2}\: \* \H(0)\, \* \zeta_4\,
   \nn \\[0mm] & & \mbox{} \vphantom{\Big(}
              + 32\, \* \H(0)\, \* \zeta_3\,
              + 3\, \* \H(0)\, \* \zeta_2\,
              + 30\, \* \zeta_3\, \* \zeta_2\,
              + \frct{373}{18}\: \* \zeta_2\,
              - \frct{125}{3}\: \* \zeta_3\,
              - 13\, \* \zeta_4\,
              - 38\, \* \zeta_5\,
              - \frct{850}{9}\,
\Big\}
\eea
where our normalization of the colour factor is $\dabctnc = 5/18$ in QCD;
for use with third-order results note the discussion below Eq.~(30) in 
Ref.~\cite{mvvDPs}.
Finally the common leading large-$\nf$ contribution to the N$^3$LO evolution 
of all three types of quark distributions in Eq.~(\ref{qns}) reads
\bea
\label{pns3nf3}
\lefteqn{ \Pns{3}(x)\Big|_{\nft} \;=\;
       \frct{32}{9}\: \* \colourcolour{\cf} \, \*  \Big\{ \pqq{x}\, \* \Big( 
              - \frct{1}{6}\: \* \Hhh(0,0,0)\,
              - \frct{5}{18}\: \* \Hh(0,0)\,
              + \frct{1}{18}\: \* \H(0)\,
              + \frct{1}{3}\: \* \zeta_3\,
              - \frct{1}{18}\Big)
} \hspace*{3cm}
   \nn \\[-0.5mm] & & \mbox{ } \vphantom{\Big(}
          + x\, \* \Big(\frct{1}{3}\: \* \Hh(0,0)
              + \frct{13}{18}\: \* \H(0)\,
              + \frct{1}{6}\Big)\,
              - \frct{1}{3}\: \* \Hh(0,0)\,
              - \frct{13}{18}\: \* \H(0)\,
              - \frct{1}{6}
   \nn \\[0.5mm] & & \mbox{} \vphantom{\Big(}
          \, + \, \delta(1-x)\, \* \Big(  
              - \frct{131}{288}
              + \frct{1}{9}\: \* \zeta_2\,
              + \frct{19}{18}\: \* \zeta_3\,
              - \frct{1}{3}\: \* \zeta_4\,
              \Big)\Big\}
\:\: .
\eea
Also the large-$x$ limit is the same for the three non-singlet splitting 
functions. It is given by 
\beq
\label{xto1}
  P_{\,\rm ns}^{\,(n-1)\pm,\rm v}(x) \:\: = \:\;
        \frac{A_n}{(1-x)_+}
  \,+\, B_n \, \delta \x1
  \,+\, C_n \, \ln(1-x)
  \,+\, D_n + {\cal O} \left( \x1 \ln^{\,\ell\!} \x1 \right)
\eeq
in terms of the same constants as in Eq.~(\ref{ntoinf}), i.e., the
$n_{\!f}^{\:a\,>1}$ contributions to $A_4$ and $C_4$ have been given in
Eqs.~(\ref{A4nf23}) and (\ref{C4nf2}). The coefficients $B_4$ can be read 
of from Eqs.~(\ref{pns3Anf2}), (\ref{pns3Bnf2}) and (\ref{pns3nf3}).
The difference between $P^{\,-}_{\rm ns}$ and $P^{\,+}_{\rm ns}$
and the splitting function (\ref{PnsS}) are suppressed by two powers
of $\x1$ with respect to the leading term in Eq.~(\ref{xto1}). 

The non-singlet splitting functions include double-logarithmic small-$x$
contributions up to $\ln^{\,2\ell\!} x$ at N$^\ell$LO. The coefficients of
these leading-logarithmic (LL) parts of $P_{\rm ns}^{\,\pm}$ have long 
been known to all orders \cite{KL82,BV95}; the presence of a $\ln^{\,4\!} x$
term in $P_{\rm ns}^{\:\rm s}$ at NNLO had not been predicted before the
three-loop calculation in Ref.~\cite{mvvPns}. 
Contributions where $k$ powers of $(C_A,\:C_F)$ in the colour factor are 
replaced by $n_{\!f}^{\:k}$ are suppressed by $k$ powers of $\ln x$ 
relative to the overall leading logarithms. Hence we expect terms up to 
$\ln^{\,4} x$ and $\ln^{\,5} x$, respectively, in $P^{\,(3)\pm}_{\rm ns}$
and $P^{\,(3)\rm s}_{\rm ns}$ at $\nfs$. Indeed we find
\bea
	P_{\rm ns}^{\,(3)+}\big|_{\nfs} &\!=\!&
\label{Pns3Pxto0}
  \: \ln^{4} x \, \*  \Big(\,
         \frct{4}{9}\: \* \cfs\,
         \Big)\,
  + \ln^{3} x \, \*  \Big(\,
	     \frct{152}{27}\: \* \cfs\,
	   + \frct{44}{27}\: \* \cf \* \ca\,
	          \Big)\,
\nn \\   & & \mbox{\hspn}
       + \ln^{2} x \, \*  \Big(\,
	     \frct{16}{81}\: \* [134 + 9\, \* \zeta_2]\, \* \cfs\,
	   + \frct{4}{27}\: \* [161 - 36\, \* \zeta_2]\, \* \cf \* \ca\,
	          \Big)\,
		\,
\nn \\   & & \mbox{\hspn}
       + \ln x \, \*  \Big(\,
	     \frct{8}{81}\: \* [967 + 72\, \* \zeta_2]\, \* \cfs\,
	   + \frct{1}{81}\: \* [7561 - 2736\, \* \zeta_2 + 864\, \* \zeta_3]\, 
           \* \cf \* \ca\,
		 \Big)
\:\: , \quad
\\[3mm]
\label{Pns3Mxto0}
	P_{\rm ns}^{\,(3)-}\big|_{\nfs} &\!=\!&
  \:  \ln^{4} x \, \*  \Big(\,
		 \frct{4}{9}\: \* \cf \* \ca\,
       - \frct{4}{9}\: \* \cfs\,
	     \Big)\,
  + \ln^{3} x \, \*  \Big(\,
	     \frct{692}{81}\: \* \cf\, \* \ca\,
	   - \frct{664}{81}\: \* \cfs\,
	     \Big)\,
\nn \\   & & \mbox{\hspn}
       + \ln^{2} x \, \*  \Big(\,
		  \frct{4}{81}\: \* [1081 - 36\, \* \zeta_2]\, \* \cf \* \ca\,
        - \frct{16}{27}\: \* [55 + 9\, \* \zeta_2]\, \* \cfs\,
		          \Big)\,
\\   & & \mbox{\hspn}
       + \ln x \, \*  \Big(\,
		  \frct{1}{27}\: \* [4131 - 304\, \* \zeta_2 + 384\, \* \zeta_3]\, 
            \* \cf \* \ca\,
	    - \frct{8}{81}\: \* [241 + 384\, \* \zeta_2 + 72\, \* \zeta_3]\, 
            \* \cfs\, \Big)
\nn \qquad
\eea
and 
\bea
\label{Pns3Sxto0}
  P_{\rm ns}^{\,(3)\rm s}\big|_{\nfs\,\dabctnc} &\!=\!\!& \mbox{}
    - \, \frct{64}{45}\: \* \ln^{5} x 
  \,-\, \frct{64}{3}\: \* \ln^{4} x 
  \,-\, \frct{128}{9}\: \* ( 3 - \z2 ) \, \* \ln^{3} x
\\   & &  \mbox{} 
    -\, \frct{256}{3}\: \* ( 14 - 9\,\* \z2 + 3\,\* \z3 )\, \* \ln^{2} x
  \,-\, \frct{64}{3}\: \* ( 138 + 26\,\* \z2 - 64\,\* \z3 + 5\,\* \z4 )
        \, \* \ln x
\qquad \nn
\eea
up to constants and terms vanishing for $x \ra 0$.
The corresponding limit of Eq.~(\ref{pns3nf3}) reads
\beq
\label{Pns3nf3to0}
  P_{\rm ns}^{\,(3)}\big|_{\cf \nft} \;=\;
       - \,\frct{8}{81}\: \* \ln^{3} x 
       - \,\frct{88}{81}\: \* \ln^{2} x 
       - \,\frct{64}{27}\: \* \ln x 
       + \,{\cal O}(1)
\:\: .
\eeq
 
The analytic structure of the LL resummations is very different for
$P_{\rm ns}^{\,+}$ and $P_{\rm ns}^{\,-}$ with \cite{KL82,BV95} 
\beq 
\label{PnsPto0LL}
  P_{\rm ns,LL}^{\,+} (N, a_{\rm s}) \;=\; \frct{N}{2}\:
  \Big\{ 1 - 
    \Big( 1 - \frct{8\:\! a_{\rm s} C_F}{N^2} 
    \Big)^{\!-1/2\,} 
  \Big\}
\eeq
and
\beq
\label{PnsMto0LL}
  P_{\rm ns,LL}^{\,-} (N, \ars) \;=\; \frct{N}{2}\:
  \Big\{ 1 - 
    \Big( 1 - \frct{8\:\! \ars C_F}{N^2}
      \Big[ 1 - \frct{8\:\! \ars n_c}{N} \: \frct{d}{dN} \,
        \ln \Big ( e^{\:z^2/4} \, D_{-1/[2n_c^2]}(z) \Big ) 
      \Big] 
    \Big)^{\!-1/2\,}
  \Big\}
\eeq
where $z = N (2\:\! \ars N_c)^{-1/2}$, and $D_p(z)$ denotes a parabolic 
cylinder function~\cite{Gradsht}. The expansion of Eq.~(\ref{PnsMto0LL}) 
in powers of $\ars$ is an asymptotic expansion, in contrast to 
Eq.~(\ref{PnsPto0LL}). The difference between the two expansions vanishes
in the large-$n_c$ limit.
 
An extension of these resummations to next-to-leading logarithmic (NLL) 
accuracy and beyond is known so far only for the former case 
 --- for the $x^{\,2n} \ln^{\,\ell\!} x$ terms at $n \geq 0$; 
 for the $x^{\,2n+1} \ln^{\,\ell\!} x$ terms the roles of $P_{\rm ns}^{\,+}$
 and $P_{\rm ns}^{\,-}$ are interchanged in this respect.
A determination of the N$^\ell$LL terms on the basis of N$^\ell$LO information 
is possible from the $D$-dimensional structure of the unfactorized expressions, 
analogous to the case of the final-state splitting functions and coefficient 
functions in semi-inclusive annihilation \cite{AV2011,KVY2012}.
The first term in Eq.~(\ref{Pns3Mxto0}), an overall NNLL contribution, agrees 
with the result in Eq.~(4.6) of Ref.~\cite{avLL2012} after $\als$-expansion and 
Mellin inversion. For details and results on the singlet cases
and coefficient functions see Ref.~\cite{DKVprp}.

A generalization of the equation underlying Eq.~(\ref{PnsPto0LL}) to all powers
of $\,\ln x\,$, i.e., the terms with $\,1/N^{\,a \,>1\,}$ in the expansion about 
$N=0$, has been suggested in Ref.~\cite{VelizDL} as
\bea
\label{Pnsto0SL}
  P_{\rm ns}^{\,+}(N,\ars) \,\left( P_{\rm ns}^{\,+}(N,\ars) 
  - N + \beta(\ars) / \ars \right) &\!=\!& O(1) 
\nn \\[1mm]
  \mbox{ up to terms with $\;\;\z2 (\ca -2\:\! \cf )\,$,} & & 
\eea
where $\beta(\ars) = - \beta_0 \,\ar(2) - \beta_1 \:\!\ar(3) - \,\ldots$ with 
$\beta_0 = 11/3\: C_A - 2/3\: \nf$ is the beta function of QCD; the terms 
including $\beta_2$ \cite{beta2a,beta2b} enter the four-loop evaluation of 
Eq.~(\ref{Pnsto0SL}). This evaluation indeed reproduces Eq.~(\ref{Pns3Pxto0}) 
except for the $\z2 (\ca -2\,\cf)$ contributions --- note that there are typos
in Eq.~(25) and (26) of Ref.~\cite{VelizDL} --- as well as the corresponding
terms up to overall NNLL accuracy resulting from Eq.~(4.6) of 
Ref.~\cite{avLL2012}.

\begin{figure}[p]
\vspace{-4mm}
\centerline{\epsfig{file=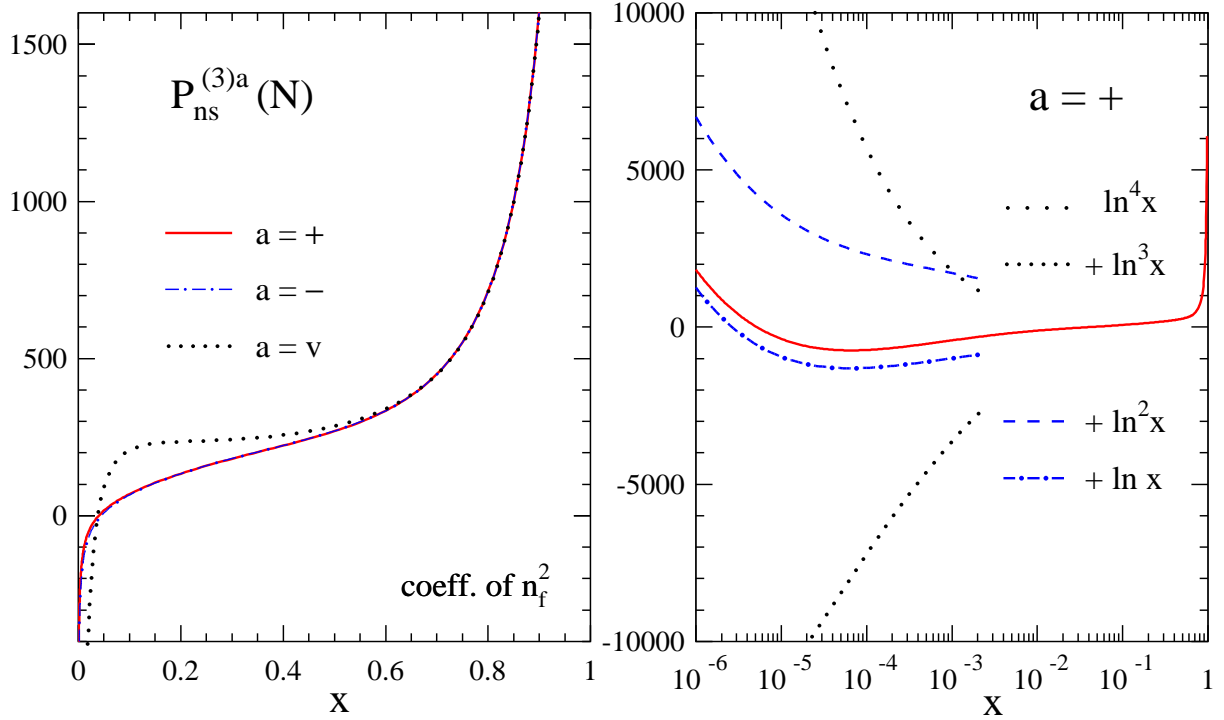,width=16.0cm,angle=0}}
\vspace{-2mm}
\caption{ \label{fig:pns1} \small
 Left: the $\nfs$ parts of the four-loop splitting functions for the evolution 
 of the combinations (\ref{qns}) of quark and anti-quark distributions given by
 Eqs.~(\ref{Pns3dec}) -- Eqs.~(\ref{pns3Snf2}).
 Right: the small-$x$ behaviour of this contribution to $P_{\rm ns}^{\:+}$,
 compared to its successive approximations by the small-$x$ logarithms in 
 Eq.~(\ref{Pns3Pxto0}).}
\vspace{-1mm}
\end{figure}

\begin{figure}[p]
\vspace{-2mm}
\centerline{\epsfig{file=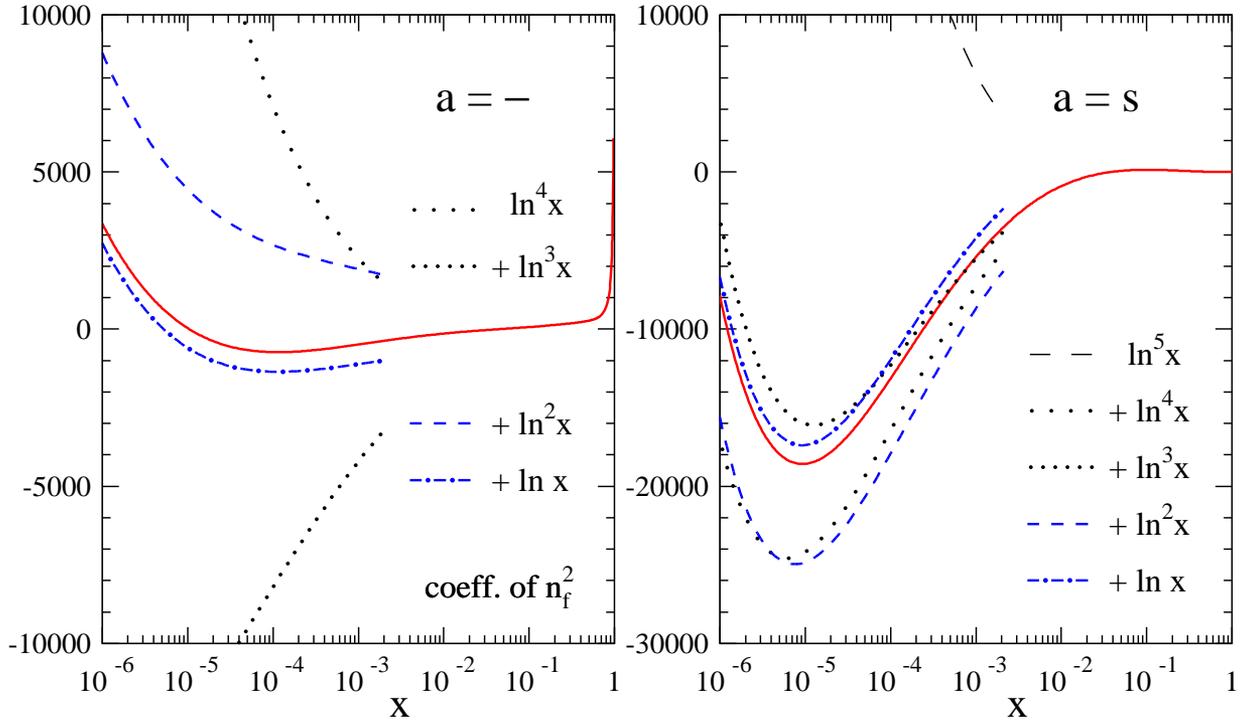,width=16.0cm,angle=0}}
\vspace{-2mm}
\caption{ \label{fig:pns2} \small
 As the right panel of Figure~\ref{fig:pns1}, but for the splitting functions
 $P_{\rm ns}^{\:-}$ (left) and $P_{\rm ns}^{\:\rm s}$ (right). 
 Due to Eq.~(\ref{Pexp}) all numbers have to be divided by $(4\:\! \pi)^4 
 \simeq 25000\,$ for an expansion in powers of $\als$.} 
\vspace{-1mm}
\end{figure}

The $\nfs$ contributions to the three non-singlet splitting functions are 
illustrated In Figs.~\ref{fig:pns1} and \ref{fig:pns2} on linear and 
logarithmic scales in $x$. The latter have been extended to $x = 10^{\,-6}$ in 
order to include the onset of the steep small-$x$ rise of all these functions.
The difference (\ref{pns3dBnf2}) between the $\nfs$ parts of 
$P_{\rm ns}^{\,(3)-}$ and $P_{\rm ns}^{\,(3)+}$ is numerically irrelevant 
except at very small $x$. 
The $\nfs\, d^{abc\,}d_{abc}/n_c$ difference between $P_{\rm ns}^{\,(3)\rm v}$
and $P_{\rm ns}^{\,(3)-}$, on the other hand, is non-negligible up to 
$x \simeq 0.5$.

At asymptotically small values of $x$, the behaviour of these functions 
is given by their respective leading $\ln^{\,4}x$ and $\ln^{\,5}x$ logarithms 
in Eqs.~(\ref{Pns3Pxto0}) -- (\ref{Pns3Sxto0}). As shown in the figures, 
though, the onset of the resulting steep rise towards $x = 0$ is delayed to 
$x \approx 10^{\,-5}$ by the effect of the non-leading logarithms. In fact,
even at the lowest $x$-values shown here a relevant approximation for
$P_{\rm ns}^{\,+}$ and $P_{\rm ns}^{\,-}$ is obtained only if all $\ln x$ 
terms are taken into account. 
The situation is more favourable for $P_{\rm ns}^{\,\rm s}$ but, unlike for 
the three-loop contribution~\cite{mvvPns} to this function, also here the 
leading logarithmic result is totally different from the actual function at
all physically sensible values of $x$.


\pagebreak

The $x$-space splitting functions corresponding to the flavour-singlet anomalous
dimensions in Eqs.~(\ref{gpsnf3}) -- (\ref{gggnf3}) are given by
\bea
\label{pps3nf3}
\lefteqn{ \left. \Pqqps{3}(x)\right|_{\nft} \;=\;
       \frct{32}{9}\: \* \colourcolour{\cf} \, \*  
       \Big\{ \frct{1}{x}\, \* \Big(
                \frct{8}{3}\: \* \Hhh(1,1,1)
              + \frct{4}{9}\: \* \H(1)\,
              - 4\, \* \zeta_3\,
              + \frct{2}{3}\Big)\,
          + (1+x)\, \* \Big( - \Hhhh(0,0,0,0)\,
              + 4\, \* \Hhh(2,1,1)\,
}
   \nn \\[-1mm] & & \mbox{} \vphantom{\Big(}
              - 4\, \* \Hh(3,1)\,
              + 2\, \* \H(4)\,
              - \frct{29}{6}\: \* \Hhh(0,0,0)\,
              + \frct{29}{3}\: \* \H(3)\,
              - \frct{73}{9}\: \* \Hh(0,0)\,
              - 2\, \* \Hh(0,0)\, \* \zeta_2\,
              - 4\, \* \H(0)\, \* \zeta_3\,
              - \frct{29}{3}\: \* \H(0)\, \* \zeta_2\,
   \nn \\[0mm] & & \mbox{} \vphantom{\Big(}
              - 5\, \* \zeta_4\Big)\,
          + x\, \* \Big( - 2\, \* \Hhh(1,1,1)\,
              - \frct{14}{3}\: \* \Hh(2,1)\,
              + 14\, \* \Hh(1,1)\,
              + \frct{2}{9}\: \* \H(2)\,
              - \frct{11}{9}\: \* \H(0)\,
              - \frct{166}{9}\: \* \H(1)\,
              - \frct{2}{9}\: \* \zeta_2\,
   \nn \\[0.5mm] & & \mbox{} \vphantom{\Big(}
              - 2\, \* \zeta_3\,
              + \frct{38}{9}\Big)\,
          + x^{2}\, \* \Big( 
              - \frct{8}{3}\: \* \Hhh(1,1,1)\,
              + \frct{8}{3}\: \* \Hh(2,1)\,
              - 4\, \* \Hh(1,1)\,
              - \frct{76}{9}\: \* \H(2)\,
              + \frct{64}{9}\: \* \H(0)\,
              + \frct{68}{9}\: \* \H(1)\,
   \nn \\[0.5mm] & & \mbox{} \vphantom{\Big(}
              + \frct{76}{9}\: \* \zeta_2\,
              + \frct{4}{3}\: \* \zeta_3\,
              - \frct{14}{9}\Big)\,
              + 2\, \* \Hhh(1,1,1)
              - \frct{26}{3}\: \* \Hh(2,1)\,
              - 10\, \* \Hh(1,1)\,
              + \frct{98}{9}\: \* \H(2)\,
              - \frct{59}{9}\: \* \H(0)\,
              + \frct{94}{9}\: \* \H(1)\,
   \nn \\[0.5mm] & & \mbox{} \vphantom{\Big(}
              - \frct{98}{9}\: \* \zeta_2\,
              - 4\, \* \zeta_3\,
              - \frct{10}{3}\Big\}
\;\; ,
\eea
\bea
\label{pqg3nf3}
\lefteqn{ \left. \Pqg{3}(x)\right|_{\nft} \;=\;
       \frct{32}{9}\: \* \colourcolour{\cf} \, \* 
       \Big\{ \frct{1}{x}\, \* \Big(
                \frct{8}{9}\: \* \Hhh(1,0,0)
              - \frct{8}{9}\: \* \Hhh(1,1,0)\,
              - \frct{8}{9}\: \* \Hhh(1,1,1)\,
              - \frct{8}{9}\: \* \Hh(1,2)\,
              - \frct{92}{27}\: \* \Hh(1,0)\,
              + \frct{4}{9}\: \* \Hh(1,1)\,
}
   \nn \\[-0.5mm] & & \mbox{} \vphantom{\Big(}
              + \frct{284}{81}\: \* \H(1)\,
              + \frct{8}{9}\: \* \H(1)\, \* \zeta_2\,
              - \frct{16}{3}\: \* \zeta_3\,
              + \frct{136}{81}\Big)\,
          + (1-2\, \* x)\, \* \Big( - 2\, \* \Hhh(3,0,0)\,
              + 2\, \* \Hhh(3,1,0)\,
              + 2\, \* \Hhh(3,1,1)\,
   \nn \\[0mm] & & \mbox{} \vphantom{\Big(}
              + 2\, \* \Hh(3,2)\,
              - 8\, \* \Hh(4,0)\,
              - 10\, \* \H(5)\,
              + \Hhhh(1,0,0,0)\,
              + \frct{1}{3}\: \* \Hhhh(1,1,1,1)\,
              + 10\, \* \Hhh(0,0,0)\, \* \zeta_2\,
              - 2\, \* \H(3)\, \* \zeta_2\,
   \nn \\[0mm] & & \mbox{} \vphantom{\Big(}
              + 6\, \* \Hh(0,0)\, \* \zeta_3\,
              - 4\, \* \H(0)\, \* \zeta_4\,
              - 2\, \* \H(1)\, \* \zeta_3\,
              - 2\, \* \zeta_3\, \* \zeta_2\,
              - 2\, \* \zeta_5\Big)\,
          + x\, \* \Big( 
              - \frct{163}{3}\: \* \Hhhh(0,0,0,0)\,
              - \frct{2}{3}\: \* \Hhh(2,1,1)\,
   \nn \\[0mm] & & \mbox{} \vphantom{\Big(}
              + \frct{8}{3}\: \* \Hh(3,0)\,
              - \frct{16}{3}\: \* \Hh(3,1)\,
              + 16\, \* \H(4)\,
              - \frct{538}{9}\: \* \Hhh(0,0,0)\,
              + \frct{109}{9}\: \* \Hhh(1,0,0)\,
              - 15\, \* \Hhh(1,1,0)\,
              - \frct{121}{9}\: \* \Hhh(1,1,1)\,
   \nn \\[0mm] & & \mbox{} \vphantom{\Big(}
              - 15\, \* \Hh(1,2)\,
              - 32\, \* \Hh(2,0)\,
              - \frct{130}{9}\: \* \Hh(2,1)\,
              - \frct{265}{9}\: \* \H(3)\,
              + \frct{341}{36}\: \* \Hh(0,0)\,
              - 16\, \* \Hh(0,0)\, \* \zeta_2\,
              - \frct{346}{27}\: \* \Hh(1,0)\,
   \nn \\[0mm] & & \mbox{} \vphantom{\Big(}
              - \frct{2029}{54}\: \* \Hh(1,1)\,
              - \frct{1262}{27}\: \* \H(2)\,
              + \frct{1426}{27}\: \* \H(0)\,
              - \frct{10}{3}\: \* \H(0)\, \* \zeta_3\,
              + \frct{265}{9}\: \* \H(0)\, \* \zeta_2\,
              + \frct{323}{27}\: \* \H(1)\,
              + 15\, \* \H(1)\, \* \zeta_2\,
   \nn \\[0mm] & & \mbox{} \vphantom{\Big(}
              + \frct{1262}{27}\: \* \zeta_2\,
              - \frct{973}{9}\: \* \zeta_3\,
              - 6\, \* \zeta_4\,
              - \frct{31627}{324}\Big)\,
          + x^{2}\, \* \Big(10\, \* \Hhhh(0,0,0,0)
              + 2\, \* \Hhhh(1,0,0,0)\,
              + \frct{2}{3}\: \* \Hhhh(1,1,1,1)\,
   \nn \\[0mm] & & \mbox{} \vphantom{\Big(}
              + \frct{8}{3}\: \* \Hhh(2,0,0)\,
              - \frct{8}{3}\: \* \Hhh(2,1,0)\,
              - 2\, \* \Hhh(2,1,1)\,
              - \frct{8}{3}\: \* \Hh(2,2)\,
              + \frct{40}{3}\: \* \Hh(3,0)\,
              + \frct{8}{3}\: \* \Hh(3,1)\,
              + \frct{40}{3}\: \* \H(4)\,
   \nn \\[0mm] & & \mbox{} \vphantom{\Big(}
              - \frct{128}{3}\: \* \Hhh(0,0,0)\,
              - 4\, \* \Hhh(1,0,0)\,
              + \frct{62}{9}\: \* \Hhh(1,1,0)\,
              + \frct{16}{3}\: \* \Hhh(1,1,1)\,
              + \frct{62}{9}\: \* \Hh(1,2)\,
              - \frct{82}{3}\: \* \Hh(2,0)\,
              - \frct{28}{3}\: \* \Hh(2,1)\,
   \nn \\[0mm] & & \mbox{} \vphantom{\Big(}
              - \frct{158}{9}\: \* \H(3)\,
              + \frct{359}{9}\: \* \Hh(0,0)\,
              - \frct{40}{3}\: \* \Hh(0,0)\, \* \zeta_2\,
              + \frct{151}{3}\: \* \Hh(1,0)\,
              + \frct{785}{27}\: \* \Hh(1,1)\,
              + \frct{547}{27}\: \* \H(2)\,
              + \frct{8}{3}\: \* \H(2)\, \* \zeta_2\,
   \nn \\[0mm] & & \mbox{} \vphantom{\Big(}
              + \frct{9425}{162}\: \* \H(0)\,
              - \frct{28}{3}\: \* \H(0)\, \* \zeta_3\,
              + \frct{158}{9}\: \* \H(0)\, \* \zeta_2\,
              + \frct{7547}{162}\: \* \H(1)\,
              - 4\, \* \H(1)\, \* \zeta_3\,
              - \frct{62}{9}\: \* \H(1)\, \* \zeta_2\,
              - \frct{547}{27}\: \* \zeta_2\,
   \nn \\[0mm] & & \mbox{} \vphantom{\Big(}
              + \frct{122}{9}\: \* \zeta_3\,
              + 12\, \* \zeta_4\,
              + \frct{821}{324}\Big)\,
              + \frct{139}{6}\: \* \Hhhh(0,0,0,0)
              - 3\, \* \Hhh(2,0,0)\,
              + 3\, \* \Hhh(2,1,0)\,
              + \frct{10}{3}\: \* \Hhh(2,1,1)\,
              + 3\, \* \Hh(2,2)\,
   \nn \\[0mm] & & \mbox{} \vphantom{\Big(}
              - \frct{40}{3}\: \* \Hh(3,0)\,
              - \frct{4}{3}\: \* \Hh(3,1)\,
              - 15\, \* \H(4)\,
              + \frct{965}{36}\: \* \Hhh(0,0,0)\,
              - \frct{71}{9}\: \* \Hhh(1,0,0)\,
              + 9\, \* \Hhh(1,1,0)\,
              + \frct{71}{9}\: \* \Hhh(1,1,1)\,
   \nn \\[0mm] & & \mbox{} \vphantom{\Big(}
              + 9\, \* \Hh(1,2)\,
              - 33\, \* \Hh(2,0)\,
              + \frct{17}{9}\: \* \Hh(2,1)\,
              - \frct{379}{9}\: \* \H(3)\,
              + \frct{2473}{36}\: \* \Hh(0,0)\,
              + 15\, \* \Hh(0,0)\, \* \zeta_2\,
              - \frct{952}{27}\: \* \Hh(1,0)\,
   \nn \\[0mm] & & \mbox{} \vphantom{\Big(}
              + \frct{232}{27}\: \* \Hh(1,1)\,
              - \frct{1415}{27}\: \* \H(2)\,
              - 3\, \* \H(2)\, \* \zeta_2\,
              + \frct{2104}{27}\: \* \H(0)\,
              + \frct{20}{3}\: \* \H(0)\, \* \zeta_3\,
              + \frct{379}{9}\: \* \H(0)\, \* \zeta_2\,
              - \frct{1640}{27}\: \* \H(1)\,
   \nn \\[0mm] & & \mbox{} \vphantom{\Big(}
              - 9\, \* \H(1)\, \* \zeta_2\,
              + \frct{1415}{27}\: \* \zeta_2\,
              + \frct{499}{18}\: \* \zeta_3\,
              - 10\, \* \zeta_4\,
              + \frct{58277}{648}\Big\}
\nn \\[1mm] & & \mbox{} \vphantom{\Big(} \hspace{-5mm}         
  +\frct{32}{9}\: \* \colourcolour{\ca} \, \*  
   \Big\{ \frct{1}{x}\, \* \Big(
                \frct{13}{27}\: \* \Hh(1,0)
              - \frct{13}{27}\: \* \Hh(1,1)\,
              - \frct{47}{81}\: \* \H(1)\,
              - \frct{4}{3}\: \* \zeta_3\,
              - \frct{14}{81}\Big)\,
          + (1-2\, \* x)\, \* \Big(\Hhhh(1,0,0,0)
   \nn \\[0mm] & & \mbox{} \vphantom{\Big(}
              + \frct{1}{3}\: \* \Hhhh(1,1,0,0)\,
              - \frct{1}{3}\: \* \Hhhh(1,1,1,0)\,
              - \frct{1}{3}\: \* \Hhhh(1,1,1,1)\,
              + \frct{1}{3}\: \* \Hhh(1,1,2)\,
              + \frct{1}{3}\: \* \Hhh(1,2,0)\,
              + \frct{1}{3}\: \* \Hhh(1,2,1)\,
              - \frct{1}{3}\: \* \Hh(1,3)\,
   \nn \\[0mm] & & \mbox{} \vphantom{\Big(}
              - \frct{2}{3}\: \* \Hh(3,0)\,
              + \frct{2}{3}\: \* \Hh(3,1)\,
              - \frct{8}{3}\: \* \H(4)\,
              - \frct{181}{72}\: \* \H(3)\,
              + \frct{8}{3}\: \* \Hh(0,0)\, \* \zeta_2\,
              + \frct{1}{3}\: \* \Hh(1,0)\, \* \zeta_2\,
              - \frct{1}{3}\: \* \Hh(1,1)\, \* \zeta_2\,
   \nn \\[0mm] & & \mbox{} \vphantom{\Big(}
              + \frct{8}{3}\: \* \H(1)\, \* \zeta_3\Big)\,
          + x\, \* \Big( 
              - \frct{8}{3}\: \* \Hhhh(-1,0,0,0)\,
              + \frct{2}{3}\: \* \Hhhh(0,0,0,0)\,
              - \frct{1}{6}\: \* \Hh(-2,0)\,
              - \frct{28}{9}\: \* \Hhh(-1,0,0)\,
              - 8\, \* \Hhh(0,0,0)\,
   \nn \\[0mm] & & \mbox{} \vphantom{\Big(}
              - \frct{14}{9}\: \* \Hhh(1,0,0)\,
              - \frct{14}{9}\: \* \Hhh(1,1,0)\,
              - \frct{14}{9}\: \* \Hhh(1,1,1)\,
              + \frct{14}{9}\: \* \Hh(1,2)\,
              + \frct{4}{3}\: \* \Hh(2,0)\,
              - \frct{5}{4}\: \* \Hh(2,1)\,
              - \frct{5}{27}\: \* \Hh(-1,0)\,
   \nn \\[0mm] & & \mbox{} \vphantom{\Big(}
              - \frct{493}{36}\: \* \Hh(0,0)\,
              + \frct{131}{24}\: \* \Hh(1,0)\,
              - \frct{1121}{216}\: \* \Hh(1,1)\,
              - \frct{29}{6}\: \* \H(2)\,
              - \frct{2449}{216}\: \* \H(0)\,
              - \frct{28}{3}\: \* \H(0)\, \* \zeta_3\,
              - \frct{187}{36}\: \* \H(0)\, \* \zeta_2\,
   \nn \\[0mm] & & \mbox{} \vphantom{\Big(}
              - \frct{67}{6}\: \* \H(1)\,
              - \frct{14}{9}\: \* \H(1)\, \* \zeta_2\,
              + \frct{251}{54}\: \* \zeta_2\,
              - \frct{169}{36}\: \* \zeta_3\,
              - \frct{17}{3}\: \* \zeta_4\,
              - \frct{35987}{1296}\Big)\,
          + x^{2}\, \* \Big( 
              - \frct{8}{3}\: \* \Hhhh(-1,0,0,0)\,
   \nn \\[0mm] & & \mbox{} \vphantom{\Big(}
              + 2\, \* \Hhhh(1,0,0,0)\,
              + \frct{2}{3}\: \* \Hhhh(1,1,0,0)\,
              - \frct{2}{3}\: \* \Hhhh(1,1,1,0)\,
              - \frct{2}{3}\: \* \Hhhh(1,1,1,1)\,
              + \frct{2}{3}\: \* \Hhh(1,1,2)\,
              + \frct{2}{3}\: \* \Hhh(1,2,0)\,
              + \frct{2}{3}\: \* \Hhh(1,2,1)\,
   \nn \\[0mm] & & \mbox{} \vphantom{\Big(}
              - \frct{2}{3}\: \* \Hh(1,3)\,
              - \frct{28}{9}\: \* \Hhh(-1,0,0)\,
              + \frct{25}{3}\: \* \Hhh(0,0,0)\,
              + \frct{14}{9}\: \* \Hhh(1,0,0)\,
              + \frct{14}{9}\: \* \Hhh(1,1,0)\,
              + \frct{14}{9}\: \* \Hhh(1,1,1)\,
              - \frct{14}{9}\: \* \Hh(1,2)\,
   \nn \\[0mm] & & \mbox{} \vphantom{\Big(}
              + \frct{13}{9}\: \* \Hh(2,0)\,
              - \frct{13}{9}\: \* \Hh(2,1)\,
              + \frct{65}{9}\: \* \H(3)\,
              - \frct{14}{27}\: \* \Hh(-1,0)\,
              - \frct{3293}{216}\: \* \Hh(0,0)\,
              - \frct{797}{216}\: \* \Hh(1,0)\,
              + \frct{2}{3}\: \* \Hh(1,0)\, \* \zeta_2\,
   \nn \\[0mm] & & \mbox{} \vphantom{\Big(}
              + \frct{829}{216}\: \* \Hh(1,1)\,
              - \frct{2}{3}\: \* \Hh(1,1)\, \* \zeta_2\,
              - \frct{2387}{216}\: \* \H(2)\,
              + \frct{8861}{1296}\: \* \H(0)\,
              - \frct{65}{9}\: \* \H(0)\, \* \zeta_2\,
              + \frct{20549}{1296}\: \* \H(1)\,
              + \frct{16}{3}\: \* \H(1)\, \* \zeta_3\,
   \nn \\[0mm] & & \mbox{} \vphantom{\Big(}
              + \frct{14}{9}\: \* \H(1)\, \* \zeta_2\,
              + \frct{2387}{216}\: \* \zeta_2\,
              + \frct{67}{9}\: \* \zeta_3\,
              + \frct{18079}{648}\Big)\,
              - \frct{4}{3}\: \* \Hhhh(-1,0,0,0)\,
              - \Hhhh(0,0,0,0)\,
              - \frct{20}{9}\: \* \Hhh(-1,0,0)\,
   \nn \\[0mm] & & \mbox{} \vphantom{\Big(}
              + \frct{29}{18}\: \* \Hhh(0,0,0)\,
              + \frct{10}{9}\: \* \Hhh(1,0,0)\,
              + \frct{10}{9}\: \* \Hhh(1,1,0)\,
              + \frct{10}{9}\: \* \Hhh(1,1,1)\,
              - \frct{10}{9}\: \* \Hh(1,2)\,
              - \frct{5}{24}\: \* \Hh(2,0)\,
              + \frct{5}{24}\: \* \Hh(2,1)\,
   \nn \\[0mm] & & \mbox{} \vphantom{\Big(}
              - \frct{19}{27}\: \* \Hh(-1,0)\,
              - \frct{277}{216}\: \* \Hh(0,0)\,
              - \frct{13}{6}\: \* \Hh(1,0)\,
              + \frct{79}{54}\: \* \Hh(1,1)\,
              - \frct{353}{72}\: \* \H(2)\,
              + \frct{539}{216}\: \* \H(0)\,
              - \frct{4}{3}\: \* \H(0)\, \* \zeta_3\,
   \nn \\[0.5mm] & & \mbox{} \vphantom{\Big(}
              + \frct{181}{72}\: \* \H(0)\, \* \zeta_2\,
              - \frct{295}{48}\: \* \H(1)\,
              + \frct{10}{9}\: \* \H(1)\, \* \zeta_2\,
              + \frct{353}{72}\: \* \zeta_2\,
              + \frct{8}{9}\: \* \zeta_3\,
              + \frct{1}{2}\: \* \zeta_4\,
              + \frct{3341}{1296}\Big\}
\;\; ,
\eea
\bea
\label{pgq3nf3}
\lefteqn{ \left. \Pgq{3}(x)\right|_{\nft} \;=\;
       \frct{32}{9}\: \* \colourcolour{\cf} \, \*  
       \Big\{ \frct{1}{x}\, \* \Big(
                \frct{4}{3}\: \* \Hhh(1,1,1)
              - \frct{20}{9}\: \* \Hh(1,1)\,
              - \frct{4}{9}\: \* \H(1)\,
              + \frct{8}{3}\: \* \zeta_3\,
              - \frct{4}{9}\Big)\,
          + x\, \* \Big(\frct{2}{3}\: \* \Hhh(1,1,1)
}
   \nn \\[0mm] & & \mbox{} \vphantom{\Big(}
              - \frct{16}{9}\: \* \Hh(1,1)\,
              + \frct{8}{9}\: \* \H(1)\,
              + \frct{4}{3}\: \* \zeta_3\Big)\,
              - \frct{4}{3}\: \* \Hhh(1,1,1)\,
              + \frct{20}{9}\: \* \Hh(1,1)\,
              + \frct{4}{9}\: \* \H(1)\,
              - \frct{8}{3}\: \* \zeta_3\,
              + \frct{4}{9}\Big\}
\hspace{2.0cm}
\eea
and
\bea
\label{pgg3nf3}
\lefteqn{ \left. \Pgg{3}(x)\right|_{\nft} \;=\;
       \frct{32}{9}\: \* \colourcolour{\cf} \, \*   
        \Big\{ (\frct{1}{x}-x^{2})\, \* \Big( 
              - \frct{4}{3}\: \* \Hhh(1,0,0)\,
              - \frct{8}{3}\: \* \Hhh(1,1,0)\,
              - \frct{8}{3}\: \* \Hhh(1,1,1)\,
              - \frct{8}{3}\: \* \Hh(1,2)\,
              + \frct{46}{9}\: \* \Hh(1,0)\,
}
   \nn \\[0mm] & & \mbox{} \vphantom{\Big(}
              + \frct{8}{3}\: \* \H(1)\, \* \zeta_2\,
              + 4\, \* \zeta_3\Big)\,
          + \frct{1}{x}\, \* \Big(
                \frct{52}{9}\: \* \Hh(1,1)
              - \frct{142}{27}\: \* \H(1)\,
              - \frct{34}{27}\Big)\,
          + (1-x)\, \* \Big( - \Hhh(1,0,0)\,
              - 2\, \* \Hhh(1,1,0)\,
   \nn \\[0mm] & & \mbox{} \vphantom{\Big(}
              - 2\, \* \Hhh(1,1,1)\,
              - 2\, \* \Hh(1,2)\,
              - 2\, \* \Hh(1,0)\,
              + 2\, \* \H(1)\, \* \zeta_2\Big)\,
          + (1+x)\, \* \Big( - \Hhhh(0,0,0,0)\,
              - 2\, \* \Hhh(2,0,0)\,
              - 4\, \* \Hhh(2,1,0)\,
   \nn \\[0mm] & & \mbox{} \vphantom{\Big(}
              - 4\, \* \Hhh(2,1,1)\,
              - 4\, \* \Hh(2,2)\,
              - 2\, \* \Hh(3,0)\,
              - 4\, \* \Hh(3,1)\,
              - 2\, \* \H(4)\,
              + \frct{14}{3}\: \* \Hh(2,1)\,
              + 2\, \* \Hh(0,0)\, \* \zeta_2\,
              + 4\, \* \H(2)\, \* \zeta_2\,
   \nn \\[0mm] & & \mbox{} \vphantom{\Big(}
              + 8\, \* \H(0)\, \* \zeta_3\,
              - 5\, \* \zeta_4\Big)\,
          + x\, \* \Big(
                \frct{29}{6}\: \* \Hhh(0,0,0)
              + \frct{29}{3}\: \* \Hh(2,0)\,
              + \frct{29}{3}\: \* \H(3)\,
              - \frct{40}{9}\: \* \Hh(0,0)\,
              + 6\, \* \Hh(1,1)\,
              + \frct{64}{9}\: \* \H(2)\,
   \nn \\[0mm] & & \mbox{} \vphantom{\Big(}
              - \frct{11}{2}\: \* \H(0)\,
              - \frct{29}{3}\: \* \H(0)\, \* \zeta_2\,
              + \frct{5}{9}\: \* \H(1)\,
              - \frct{64}{9}\: \* \zeta_2\,
              + 2\, \* \zeta_3\,
              - \frct{43}{27}\Big)\,
          + x^{2}\, \* \Big(
                \frct{4}{3}\: \* \Hhh(0,0,0)
              + \frct{8}{3}\: \* \Hh(2,0)\,
   \nn \\[0mm] & & \mbox{} \vphantom{\Big(}
              + \frct{8}{3}\: \* \Hh(2,1)\,
              + \frct{8}{3}\: \* \H(3)\,
              - \frct{46}{9}\: \* \Hh(0,0)\,
              - \frct{16}{9}\: \* \Hh(1,1)\,
              - \frct{16}{9}\: \* \H(2)\,
              - \frct{74}{27}\: \* \H(0)\,
              - \frct{8}{3}\: \* \H(0)\, \* \zeta_2\,
              - \frct{74}{27}\: \* \H(1)\,
   \nn \\[0mm] & & \mbox{} \vphantom{\Big(}
              + \frct{16}{9}\: \* \zeta_2\,
              + \frct{58}{27}\Big)\,
              + \frct{23}{6}\: \* \Hhh(0,0,0)
              + \frct{23}{3}\: \* \Hh(2,0)\,
              + \frct{23}{3}\: \* \H(3)\,
              - \frct{58}{9}\: \* \Hh(0,0)\,
              - 10\, \* \Hh(1,1)\,
              - \frct{68}{9}\: \* \H(2)\,
   \nn \\[0mm] & & \mbox{} \vphantom{\Big(}
              + \frct{7}{2}\: \* \H(0)\,
              - \frct{23}{3}\: \* \H(0)\, \* \zeta_2\,
              + \frct{67}{9}\: \* \H(1)\,
              + \frct{68}{9}\: \* \zeta_2\,
              + 6\, \* \zeta_3\,
              + \frct{19}{27}\,
          + \frct{77}{432}\:\* \delta(1-x) \Big\}
\nn \\[3mm] & & \mbox{} \vphantom{\Big(} \hspace{-5mm}   
   +\frct{32}{9}\: \* \colourcolour{\ca} \, \*  
          \Big\{ \pgg{x}\, \* \Big(
                \frct{2}{3}\: \* \zeta_3
              - \frct{1}{9}\Big)\,
          + (\frct{1}{x}-x^{2})\, \* \Big( 
              - \frct{13}{18}\: \* \Hh(1,0)\,
              - \frct{13}{9}\: \* \Hh(1,1)\,
              + \frct{215}{216}\: \* \H(1)\,
              + \frct{8}{27}\Big)\,
   \nn \\[0mm] & & \mbox{} \vphantom{\Big(}
          + (1-x)\, \* \Big(
                \frct{11}{24}\: \* \Hh(1,0)
              + \frct{11}{12}\: \* \Hh(1,1)\,
              - \frct{7}{9}\: \* \H(1)\Big)\,
          + (1+x)\, \* \Big( 
              - \frct{1}{3}\: \* \Hhh(0,0,0)\,
              - \frct{2}{3}\: \* \Hh(2,0)\,
              - \frct{4}{3}\: \* \Hh(2,1)\,
   \nn \\[0mm] & & \mbox{} \vphantom{\Big(}
              - \frct{2}{3}\: \* \H(3)\,
              + \frct{2}{3}\: \* \H(0)\, \* \zeta_2\Big)\,
          + x\, \* \Big(
                \frct{43}{72}\: \* \Hh(0,0)
              + \frct{43}{36}\: \* \H(2)\,
              - \frct{7}{72}\: \* \H(0)\,
              - \frct{43}{36}\: \* \zeta_2\,
              + \frct{2}{3}\: \* \zeta_3\,
              + \frct{103}{432}\Big)\,
   \nn \\[0.5mm] & & \mbox{} \vphantom{\Big(}
          + x^{2}\, \* \Big(
                \frct{13}{18}\: \* \Hh(0,0)
              + \frct{13}{9}\: \* \H(2)\,
              - \frct{215}{216}\: \* \H(0)\,
              - \frct{13}{9}\: \* \zeta_2\Big)\,
              + \frct{19}{18}\: \* \Hh(0,0)
              + \frct{19}{9}\: \* \H(2)\,
              - \frct{7}{8}\: \* \H(0)\,
              - \frct{19}{9}\: \* \zeta_2\,
   \nn \\[0.5mm] & & \mbox{} \vphantom{\Big(}
              + \frct{2}{3}\: \* \zeta_3\,
              - \frct{103}{432}\,
          + \frct{5}{864}\: \* \delta(1-x) \Big\}
\eea
with
\beq
\label{pgg}
   \pgg{x} \;=\; (1-x)^{-1} + x^{\,-1} - 2 + x - x^{\,2}
\:\: .
\eeq

The pure-singlet splitting function $P_{\rm ps}(x)$ is suppressed by two 
powers of $\x1$ in the limit $x \ra 1$, hence the large-$x$ limit of 
$P_{\rm qq}(x)$ is given by Eq.~(\ref{xto1}). The same functional form
holds for the large-$x$ expansion of $P_{\rm gg}(x)$. The $\nft$
contribution to $A_{4,\rm g}$ is related to Eq.~(\ref{A4nf23}) for 
$A_4 \equiv A_{4,\rm q}$ by the Casimir scaling $C_A / C_F$. The $\nft$
part of $B_{4,\rm g}$ can be readily read off from Eq.~(\ref{pgg3nf3}).
As for the quark case in Eq.~(\ref{C4nf2}), non-vanishing contributions 
to $C_{4,\rm g}$ occur only for $n_{\!f}^{\:a \,< 3}$.

Unlike these diagonal quantities, the off-diagonal entries $P_{\rm qg}$
and $P_{\rm gq}$ in Eq.~(\ref{Sevol}) show a double-logarithmic large-$x$
enhancement, i.e., terms up to $\ln^{\,2 n\!}\x1$ contribute to 
$P_{\rm qg}^{\,(n)}(x)$ and $P_{\rm qg}^{\,(n)}(x)$. The highest three of
these have been deduced at order $\as(4)$ from the large-$x$ behaviour of
physical evolution kernels of DIS structure functions in Ref.~\cite{SMVV09} 
and verified and resummed to all orders in Ref.~\cite{ASV10}; a closed form 
of the next-to-next-next-to-leading logarithmic (N$^3$LL) terms has been
obtained in Ref.~\cite{ALPV15}.
The large-$x$ enhanced contributions to Eqs.~(\ref{pqg3nf3}) and 
(\ref{pgq3nf3}) read
\bea
    P_{\rm qg}^{\,(3)}\big|_{\nft} &\!=\!& \;
\label{Pqg3Pxto1}
  \ln^{\,4\!} \x1 \:\, \*  
     \frct{4}{81}\: \* ( \cf - \ca )
  \: + \: \ln^{\,3\!} \x1 \:\, \*  
       \frct{160}{243}\: \* ( \cf - \ca )
\nn \\[0.5mm] & & \mbox{\hspn}
  - \: \ln^{\,2\!} \x1 \, \*  \Big(\,
       \frct{16}{243}\: \* ( 10 - 9\,\* \z2 )\, \* \ca 
     - \frct{232}{243}\: \* \cf
       \Big)\,
\\[1mm] & & \mbox{\hspn}
  + \: \ln \x1 \: \*  \Big(\,
       \frct{32}{243}\: \* ( 55 + 30\,\* \z2 - 36\,\* \z3 )\, \* \ca
      - \frct{16}{243}\: \* (71 - 108\,\* \z3)\,\* \cf
       \Big)\,
  \, + \, {\cal O}(1) \qquad
\nn
\eea
and
\bea
    P_{\rm gq}^{\,(3)}\big|_{\nft} &\!=\!& \;
\label{Pgq3Pxto1} 
 \: - \: \frct{32}{81} \: \* \ln^{\,3\!} \x1 \: \* \cf   
 \: - \: \frct{256}{81} \: \* \ln^{\,2\!} \x1 \: \* \cf
 \: - \: \frct{256}{81} \: \* \ln \x1 \: \* \cf
 \; + \; {\cal O}(1)
\:\: . \quad 
\eea
The coefficient of $\ln^{\,4\!} \x1$ in Eq.~(\ref{Pqg3Pxto1}), and the
lack of a $\ln^{\,4\!} \x1$ contributions in Eq.~(\ref{Pgq3Pxto1}), 
agree with the results of Refs.~\cite{SMVV09}. 
The same holds for the power-suppressed $\x1^{\,a} \ln^{\,4\!} \x1$,
terms at all $a \geq 1$ resulting from Eqs.~(\ref{pqg3nf3}) and the 
corresponding $\x1^{\,a} \ln^{\,3\!} \x1$ coefficients of the large-$x$ 
expansions of Eqs.~(\ref{pps3nf3}) and (\ref{pgg3nf3}), as given by the
last lines of Eqs.~(5.15) and (5.19) of Ref.~\cite{SMVV09} together with
the relation (5.20) between the pure-singlet and gluon-gluon results.

Like their non-singlet counterparts, the singlet splitting functions 
receive a double-logarithmic small-$x$ enhancement of the form $\as(n) 
\ln^{\,\ell\!} x$ with $0 \leq \ell \leq 2\:\! n$.
However, the small-$x$ behaviour in the singlet case is dominated by
additional single-logarithmic $\,x^{\,-1} \ln^{\,\ell\!} x\,$ terms,
see Refs.~\cite{BFKL1,BFKL2,Jaros,CH94}. 
In~the present $\as(4) \nft$ cases, only non-logarithmic $x^{\,-1}$ terms
occur and the small-$x$ expansions read
\bea
    P_{\rm ps}^{\,(3)}\big|_{\nft} &\!=\!&
\label{Pps3xto0}
  \frct{1}{x} \:\, \*  
         \frct{64}{27}\: \* ( 1 - 6\, \* \z3 )\, \* \cf
 \; - \; \ln^{4} x \:\, \*  
         \frct{4}{27}\: \* \cf\,
 \; - \; \ln^{3} x \:\, \*  
         \frct{232}{81}\: \* \cf\,
\nn \\[0.5mm]   & & \mbox{\hspn} 
    - \; \ln^{2} x \:\, \*  
         \frct{16}{81}\: \* ( 73 + 18\, \* \z2 )\, \* \cf
 \; - \; \ln x \:\, \*  
         \frct{32}{81}\: \* ( 59 + 87\, \* \z2 + 36\, \* \z3 ) \, \* \cf 
 \; + \; {\cal O}(1)
\:\: ,
\\[4mm]
    P_{\rm qg}^{\,(3)}\big|_{\nft} &\!=\!&
\label{Pqg3xto0}
  \frct{1}{x} \:\, \* \Big\{
         \frct{256}{729}\: \* ( 17 - 54\, \* \z3 ) \, \* \cf
   \,-\: \frct{64}{729}\: \*  ( 7 + 54\, \* \z3 ) \, \* \ca
         \Big\}
\nn \\[1mm]   & & \mbox{\hspn} 
    + \; \ln^{4} x \:\, \* \Big\{ 
         \frct{278}{81}\: \* \cf
   \,-\: \frct{4}{27}\: \* \ca
         \Big\}
 \; + \; \ln^{3} x \:\, \* \Big\{
         \frct{20}{243}\: \* ( 193 + 72\, \* \z2 ) \, \* \cf
   \,+\: \frct{232}{243}\: \* \ca
         \Big\}
\nn \\[1mm]   & & \mbox{\hspn} 
    + \; \ln^{2} x \:\, \* \Big\{
         \frct{4}{27}\: \* ( 835 + 180\, \* \z2 + 72\, \* \z3 ) \, \* \cf
   \,-\: \frct{2}{243}\: \* ( 277 - 576\, \* \z2 ) \, \* \ca
         \Big\}
\nn \\[1mm]   & & \mbox{\hspn} 
    + \; \ln\, x \:\, \* \Big\{
         \frct{32}{243}\: \* ( 1988 + 1137\, \* \z2 + 180\, \* \z3 
         - 108\, \* \z4 ) \* \, \cf
\nn \\[0mm]   & & \mbox{\hspp\hspp} 
   \,+\: \frct{4}{243}\: \* ( 643 + 543\, \* \z2 - 288\, \* \z3 ) \, \* \ca
         \Big\}
 \: + \: {\cal O}(1)
\:\: ,
\\[3mm]
    P_{\rm gq}^{\,(3)}\big|_{\nft} &\!=\!&
\label{Pgq3xto0}
    - \, \frct{1}{x} \:\, \* 
         \frct{128}{81}\: \* ( 1 - 6\, \* \z3 ) \, \* \cf
 \; + \; {\cal O}(1)
\:\: ,
\\[3mm]
    P_{\rm gg}^{\,(3)}\big|_{\nft} &\!=\!&
\label{Pgg3xto0}
  \frct{1}{x} \:\, \* \Big\{
         \frct{32}{243}\: \* ( 5 + 18\, \* \z3 ) \, \* \ca
   \,-\: \frct{64}{243}\: \* ( 17 - 54\, \* \z3 ) \, \* \cf
         \Big\}
 \; - \; \ln^{4} x \:\, \* 
         \frct{4}{27}\: \* \cf
\nn \\[1mm]   & & \mbox{\hspn} 
    + \; \ln^{3} x \:\, \* \Big\{
         \frct{184}{81}\: \* \cf
   \,-\: \frct{16}{81}\: \* \ca
         \Big\}
 \; + \; \ln^{2} x \:\, \* \Big\{
         \frct{152}{81}\: \* \ca
   \,-\: \frct{32}{81}\: \* ( 35 - 9\, \* \z2 )\, \* \cf
         \Big\}
\\[1mm]   & & \mbox{\hspn} 
    + \; \ln\, x \:\, \* \Big\{
         \frct{16}{81}\: \* ( 179 - 138\, \* \z2 + 144\, \* \z3 )\, \* \cf
   \,-\: \frct{4}{81}\: \* ( 115 - 48\, \* \z2 ) \, \* \ca
         \Big\}
 \: + \: {\cal O}(1)
\:\: . \quad \nn
\eea
The coefficients of $\ln^{\,4\!}x$ in these results agree with the
results of the double-logarithmic small-$x$ resummation \cite{DKVprp}.
The pattern in Eq.~(\ref{Pgq3xto0}), no small-$x$ logarithms, is the same
as for the $\cf \nfs$ contribution to $P_{\rm gq}$ at order $\as(3)$.

The $x^{\,-1}$ terms in Eqs.~(\ref{Pqg3xto0}) and (\ref{Pgg3xto0}) show an 
interesting feature in the large-$n_c$ limit $C_F \ra \frac{1}{2}\:n_c\,$: 
the resulting coefficients of $x^{\,-1} n_c$ for $P_{\rm qg}^{\,(3)}$ and
$P_{\rm gg}^{\,(3)}$ are identical to those of $x^{\,-1}\,C_F$ for
$P_{\rm ps}^{\,(3)}$ and $P_{\rm gq}^{\,(3)}$ in Eqs.~(\ref{Pps3xto0}) and 
(\ref{Pgq3xto0}), respectively. 
For the QCD values of the colour factors, the ratio between the $x^{\,-1}$
coefficients is 2.11 for the upper-row splitting functions and 2.09 for
their lower-row counterparts; hence these ratios are between their overall 
large-$n_c$ limit of 2 and the Casimir-scaling value of 9/4.

\begin{figure}[p]
\vspace{-4mm}
\centerline{\epsfig{file=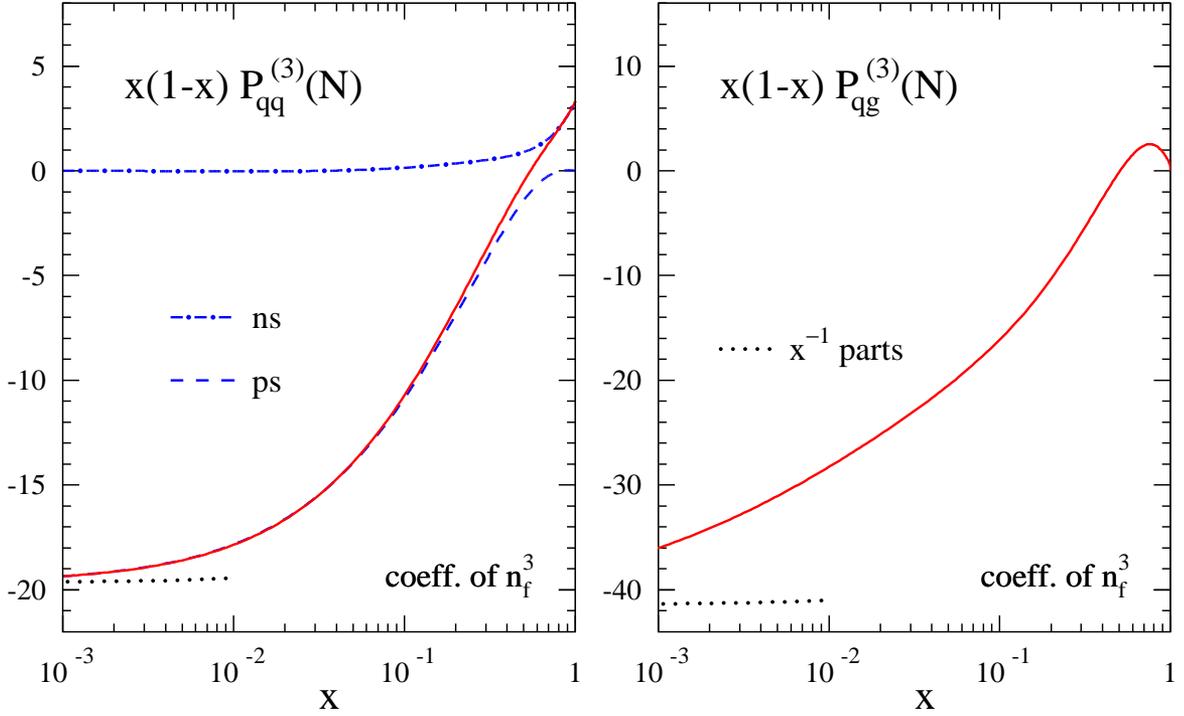,width=16.0cm,angle=0}}
\vspace{-2mm}
\caption{ \label{fig:prow1nf3} \small
 The $\nft$ parts of the `upper row' quark-quark and gluon-quark four-loop 
 splitting functions in the \MSb\ scheme, multiplied by $x\x1$ for display 
 purposes, together with their $x^{\,-1}$ leading small-$x$ terms at 
 $x < 10^{\,-2}$. For the quark-quark case also the the non-singlet and 
 pure-singlet contributions are shown.}
\vspace{-1mm}
\end{figure}

\begin{figure}[p]
\vspace{-2mm}
\centerline{\epsfig{file=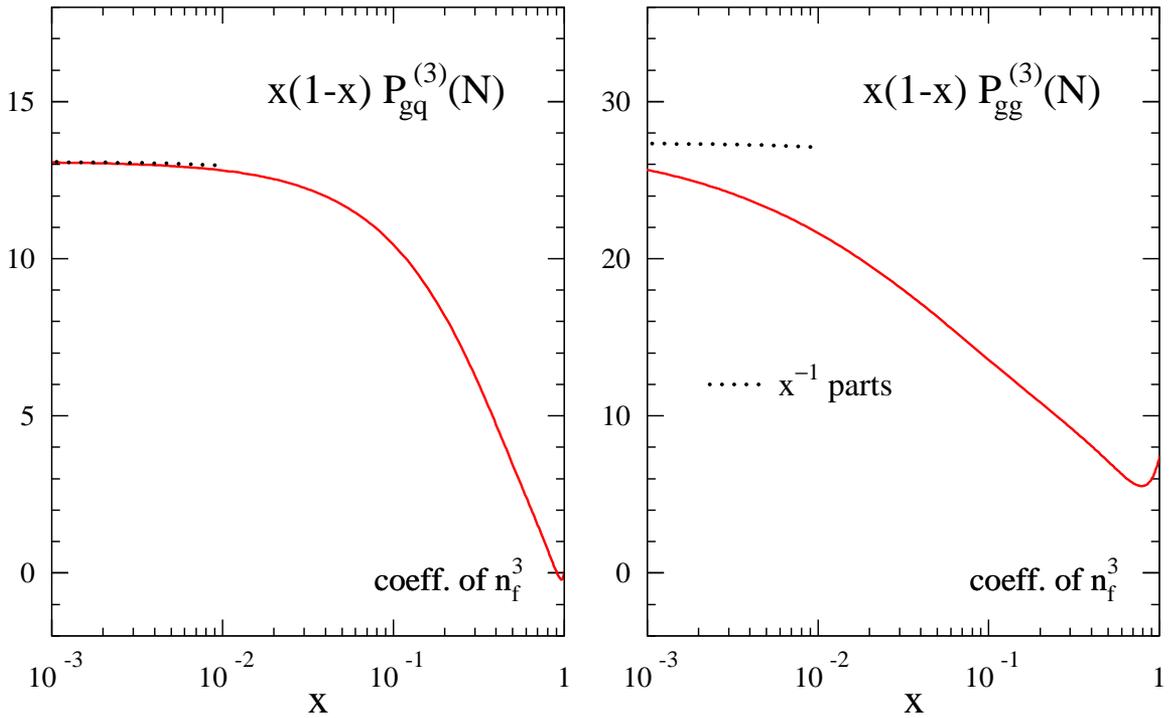,width=16.0cm,angle=0}}
\vspace{-2mm}
\caption{ \label{fig:prow2nf3} \small
 As Figure~\ref{fig:prow1nf3}, but for the `lower row' quark-gluon 
 and gluon-gluon splitting functions at N$^3$LO.}
\vspace{-1mm}
\end{figure}

The leading large-$\nf$ 4-loop contributions for the splitting function
$P_{\rm qq}(x) = P_{\rm ns}(x) + P_{\rm ps}(x)$ given by Eq.~(\ref{pns3nf3}) 
and (\ref{pps3nf3}), and those for $P_{\rm qg}(x)$, $P_{\rm gq}(x)$ and 
$P_{\rm gg}(x)$ given by Eqs.~(\ref{pqg3nf3}) -- (\ref{pgg3nf3}) are 
illustrated at $x < 1$ in Figs.~\ref{fig:prow1nf3} and \ref{fig:prow2nf3}. 
All functions have been multiplied by $x\x1$, hence their small-$x$ and 
large-$x$ limits are constants in the figures. 
For these $\ar(4)\nft$ coefficients, the pure-singlet contribution to 
$P_{\rm qq}$ remains relevant up to rather large values of $x$. 
The importance of the $\ln^{\,\ell\!}x$ small-$x$ terms is largest for 
$P_{\rm qg}$ and $P_{\rm gg}$.


\newpage
 
%
\section{Summary}
\label{sec:summ}
%

As a first step towards the determination of the N$^3$LO splitting functions 
$P_{\rm ab}^{\,(3)}(x)$ in perturbative QCD beyond the leading large-$\nf$ 
results of Refs.~\cite{LargeNf1,LargeNf2,LargeNf3}, 
we have derived the complete $\nfs$ parts of the four-loop non-singlet 
quark-quark splitting functions and all $\nft$ contributions to their 
flavour-singlet counterparts in the \MSb\ scheme.
These results have been obtained by analytically computing a fairly large
number of Mellin moments $N$ in the approach of 
Refs.~\cite{Mom3loop1,Mom3loop2,Mom3loop3} ~--~
made possible by the development of the {\sc Forcer} program 
\cite{tuLL2016,FORCER} for the computation of massless four-loop self-energy 
integrals ~--~ and a subsequent determination of the all-$N$ and all-$x$
expressions using the number-theoretical results and tools of 
Refs.~\cite{LLL,axbAlg,Calc}, a method that has been applied already 
to three-loop splitting functions in Refs.~\cite{VelizTrv,mvvDP2}.

Our results agree with Refs.~\cite{LargeNf1,LargeNf2,LargeNf3}, with the
pioneering low-$N$ non-singlet computations of 
Refs.~\cite{VelizN2,VelizN34,ChP4ns2},
and with the recent determinations of $\nf$ contributions to the four-loop 
cusp anomalous dimension~\cite{HSSS16,GHKM15,GrozinLL} which appear in our 
results as the coefficient of $\ln N$ at large $N$ or $1/(1-x)_+$ in the 
large-$x$ expansion.
We also agree with the prediction of Ref.~\cite{DMS05} for the coefficient
of $\,\ln \x1\,$ in the non-singlet cases and, in the small region of overlap, 
with the resummations of highest three small-$x$ and large-$x$ double 
logarithms in Refs.~\cite{avLL2012,DKVprp,SMVV09,ASV10}. 
Most interestingly our results are in agreement with the remarkably simple
(if incomplete -- the $\z2 (C_A - 2\:\! C_F)$ contributions are excluded)
generalization of the leading-log small-$x$ resummation \cite{KL82,BV95} for 
the quark$+$antiquark non-singlet splitting function $P_{\rm ns}^{\,+}$
to all powers of $\,\ln x\,$ proposed in Ref.~\cite{VelizDL}.

By themselves the present results are not phenomenologically useful. We hope, 
though, that it will be possible to complement them in the near future by
approximate expressions of the remaining (and numerically more important) 
contributions to the functions $P_{\rm ab}^{\,(3)}(x)$, analogous to those 
employed at NNLO~\cite{NVappr} before the results \cite{mvvPns,mvvPsg} became 
available, and hence facilitate improved N$^3$LO analyses of DIS and hard 
processes at colliders. 
One may also hope that the present results will provide useful additional
`data' for future studies of the structure of the perturbation series for
the splitting functions which, in turn, may lead to more explicit four-loop
calculations and results.

{\sc FORM} \cite{FORM3,TFORM,FORM4} files of our $N$-space expressions in terms
of harmonic sums \cite{Hsums,BKurth} and their $x$-space counterparts in terms 
of harmonic polylogarithms \cite{Hpols} can be obtained from the preprint 
server \ {\tt http://arXiv.org} by downloading the source of this article.
Furthermore they are available from the authors upon request.

%
\subsection*{Acknowledgments}
%

This work has been supported by the UK {\it Science \& Technology
Facilities Council}$\,$ (STFC) grants ST/L000431/1 and ST/K502145/1 and 
the {\it European Research Council}$\,$ (ERC) Advanced Grant 320651, 
{\it HEPGAME}. 
We also are grateful for the opportunity to use a substantial part of the 
{\tt ulgqcd} computer cluster in Liverpool which was funded by the 
STFC grant ST/H008837/1. 
In addition some computations were carried out on the Chadwick 
cluster of the University of Liverpool.


{\small
\setlength{\baselineskip}{0.4cm}

}
\end{document}